\documentclass[%
 reprint,
 superscriptaddress,
 amsmath,amssymb,
 aps,
 prl,
]{revtex4-2}

\usepackage{physics}
\usepackage{graphicx}
\usepackage{dcolumn}
\usepackage{bm}
\usepackage{booktabs} 
\usepackage[flushleft]{threeparttable}

\usepackage{xcolor}
\begin{document}

\title{Enabling Domain-Specific Atomistic Models: A Machine Learning Potential for the Solid Acid Family}

\author{Jonas H{\"a}nseroth}
\email{jonas.haenseroth@tu-ilmenau.de}
\affiliation{Theoretical Solid State Physics, Institute of Physics, Technische Universit{\"a}t Ilmenau, 98693 Ilmenau, Germany}

\author{Rose Asuka Baroness von Stackelberg}
\affiliation{Theoretical Solid State Physics, Institute of Physics, Technische Universit{\"a}t Ilmenau, 98693 Ilmenau, Germany}

\author{Christian Dre{\ss}ler}
\affiliation{Theoretical Solid State Physics, Institute of Physics, Technische Universit{\"a}t Ilmenau, 98693 Ilmenau, Germany}

\date{\today}

\begin{abstract}

Machine-learned interatomic potentials trained across the periodic table have made atomistic simulation broadly accessible, and specializing them to a single compound class is widely expected to improve accuracy. 
Yet examples remain scarce, and fewer still surpass universal models in speed or reach a higher level of electronic-structure theory.
Here we present a potential that is universal within the class of water-free solid-state hydrogen-bond network mediated proton conductors rather than across chemistry and a database of 4.4 million first-principles configurations spanning 55 materials.
It surpasses leading general-purpose potentials across this domain, recovering measured activation energies and the ordering of anion rotational dynamics that those models miss; agreement on static structure does not imply agreement on transport. 
Accuracy falls for compositions far from the training set but stays competitive for close structural relatives, and a higher level of electronic-structure theory is reached with a few hundred additional configurations per material. 
We further introduce a compact variant carrying a fifth of the parameters, faster still and yet more accurate than every general-purpose model tested. 
It sustains more than a quarter of a million atoms on a single graphics processor, placing grain boundaries and the transition into the highly conducting phase within reach, and making a quantum treatment of the protons affordable. 
Reference accuracy and accessible system size thus become largely independent, offering a template for other compound classes.

\end{abstract}

\maketitle

Fuel cell electrolytes remain confined to two regimes: hydrated polymer membranes below $100~^\circ$C and oxide-ion or proton-conducting ceramics above $600~^\circ$C.
The window between them promises faster electrode kinetics and greater tolerance towards fuel impurities at temperatures still low enough for inexpensive real-world application, yet neither class performs well with in the so-called Norby-gap~\cite{norby1999,goni12}.
Solid acids are among the most developed candidates for it: since the first demonstration of a fuel cell operating on the superprotonic phase of CsHSO$_4$~\cite{haile2001}, the compound class has been shown to sustain proton conductivities on the order of $10^{-2}$~S/cm without water acting as a charge carrier~\cite{boysen04, chisholm09, goni12}.
Practical devices have nonetheless converged on CsH$_2$PO$_4$, leaving conductivity and stability across the wider compositional space largely unmapped.

\begin{figure*}
    \centering
    \includegraphics{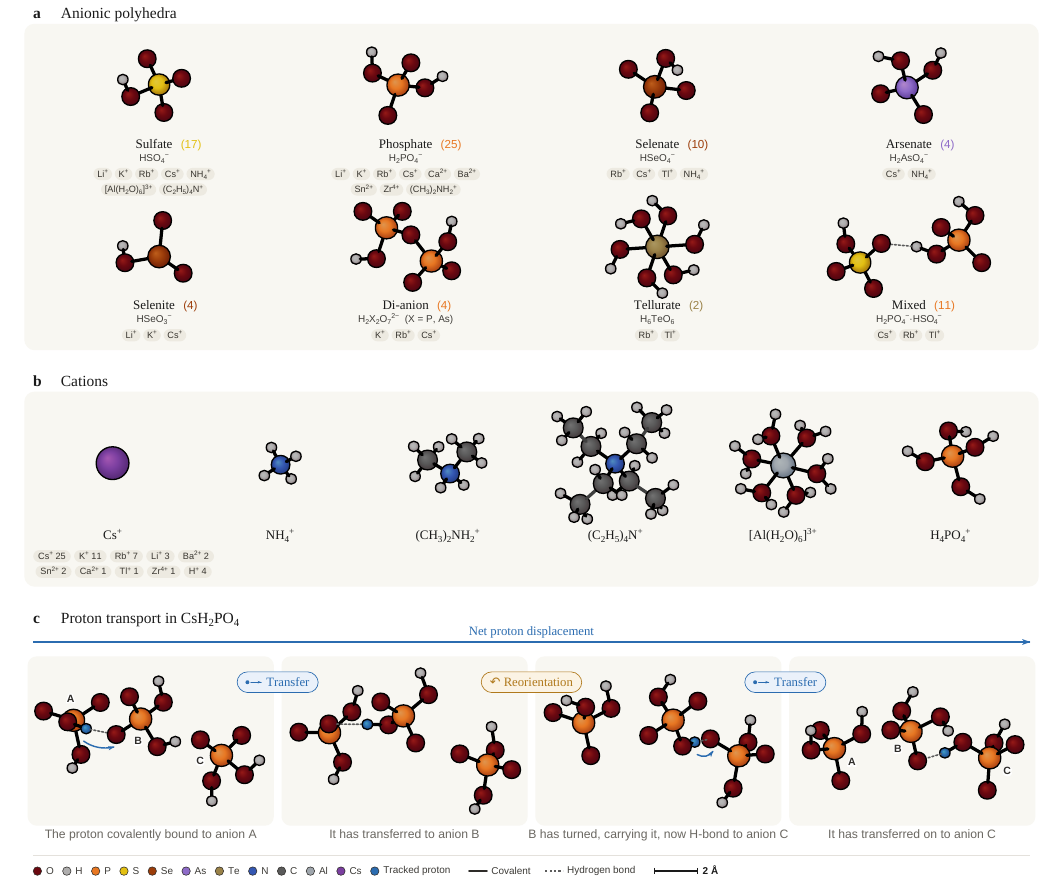}
    \caption{\textbf{Solid acid training set: chemistry and proton-transport mechanism.}
    \textbf{a} Anionic polyhedra across the $55$~materials; the number beside each name counts the materials containing that polyhedron and below are list of the cations it is found with. 
    \textbf{b} Cations, including the molecular ones.
    \textbf{c} One proton walk taken from an AIMD trajectory of CsH$_2$PO$_4$ at $510$~K: the interplay of proton transfer between different anions and subsequent polyhedron reorientation.
    All structures are AIMD snapshots~\cite{goodfellow2026xyzgraph}, without showing the cesium atoms; every molecule in the figure is drawn at the same scale. 
    Scale bar, 2~{\AA}.
    }
    \label{fig:dataset}
\end{figure*}

Solid acids are oxoanion salts of general composition AH$_y$XO$_z$, in which hydrogen bonds link neighboring oxoanions into an extended network.
The A site is usually an alkali or alkaline-earth cation, although charged organic moieties have been reported as well~\cite{mp1200282_cond}, and the oxoanion is typically tetrahedral ($z=4$) with X~$=$~P, S, Se or As.
Above a compound-specific transition temperature the oxoanion sublattice becomes orientationally disordered and proton transport proceeds by a two-step Grotthuss-type mechanism: a proton hops between the oxygen atoms of two neighboring anions, and the subsequent reorientation of the anion re-establishes a hydrogen-bond configuration for the next transfer (see Fig.~\ref{fig:dataset}c); further mechanisms have also been reported~\cite{wood2007cshso4aimd, lee2008csh2po4aimd}.
With no vehicular transport by water molecules involved, conduction is decoupled from hydration, removing the humidity management and the associated degradation pathways that constrain polymer electrolyte membranes.

Resolving a single transfer step means following the breaking and formation of one O--H bond in time, which requires a description of the electronic structure.
Ab initio molecular dynamics (AIMD) based on density functional theory (DFT) provides exactly this, but restricts studies to small simulation cells and trajectories of tens of picoseconds, well short of what converged transport coefficients require and far short of what high-throughput screening would demand~\cite{marx2000ab, tuckerman2002ab, grunert2025}.
Empirical non-reactive force fields invert the trade-off: they are cheap enough for the required length and time scales, but cannot describe bond formation and breaking and are therefore structurally unable to capture the central process in these materials~\cite{baranov03, struct_cdp, SEVIL20041659, imidazol_polymer, haile07, dressler2020effect, jinnouchi2022}.
Multiscale Monte-Carlo strategies bridge part of this gap by coarse-graining the proton dynamics onto a lattice parameterized from short ab initio trajectories~\cite{dressler2016, kabbe2016, kabbe2017, qaisrani2025bridging, haenseroth2025lmc}, but require a mechanistic model to be specified in advance.

Machine learning interatomic potentials (MLIPs) resolve this dilemma by learning the potential energy surface directly from first-principles reference data, retaining DFT accuracy at orders of magnitude lower cost~\cite{behler2007, bartok2010, drautz2019, friederich2021, reiser2022, batzner2022}.
The field has since converged on universal, or foundation, models trained on large and chemically heterogeneous datasets and applied zero-shot, that is, to systems for which no additional training data were provided~\cite{kovacs2023, mattersim, matbench, mace_mp, sol3r}.
Breadth, however, is bought at the price of resolution: the strong hydrogen bonds and the tight coupling between proton transfer and anion reorientation that define solid acids are precisely what broadly trained models do not yet reproduce accurately~\cite{jinnouchi2022, grunert2025}.
Material-specific fine-tuning reliably recovers this accuracy from a small amount of reference data~\cite{grunert2025, haenseroth2026amaceingtoolkit, radova2025fine, hanseroth2025optimizing, haenseroth2026dataset_size}, but each compound then demands its own DFT data, training run and validation and yields a potential confined to a single composition, so the cost of a comparative study grows linearly with the number of candidates.

Between these two extremes sit potentials that are universal within a compound class rather than across the periodic table, trading universal-model breadth for accuracy in a bounded chemical domain while remaining a single model.
The graphene-oxide potential \textsc{GO-MACE-23} illustrates both the approach and its principal constraint: force errors on structural motifs absent from the training set fell to benchmark level only once those motifs were included, whereas zero-shot application to isolated organic molecules reached energy errors below chemical accuracy, tracking the overlap of local bonding motifs rather than element coverage alone~\cite{elmachachi2024gomace, mahmoud2025mace_go}.
The reach of such a model is therefore set by the diversity of local environments its dataset spans, a property of the sampling strategy rather than of the compound class itself.
Comparable potentials exist for perovskite oxides~\cite{wu2024perovskitemlip}, ternary carbides~\cite{roberts2024ternarycarbidesmlip}, transition-metal carbides~\cite{dai2025transitionmetalcarbidesmlip} and diborides~\cite{dai2024transitionmetaldiboridesmlip}, liquid electrolytes~\cite{gong2025liquidelectrolytesmlip} and metallic systems~\cite{li2026metallicmatmlip}, but none, to the best of our knowledge, for solid acid proton conductors.

Our aim is therefore a local universal MLIP for solid acids: a model inheriting the transferability of the universal-model ansatz within a bounded chemical domain, while retaining the accuracy that has so far required material-specific fine-tuning or training from scratch~\cite{grunert2025, haenseroth2026amaceingtoolkit, haenseroth2026umlip_config_space, haenseroth2026dataset_size}.
As such a model is only as good as the configuration space it is trained on, the underlying dataset draws on two complementary sources.
The first comprises well-characterized solid acids studied extensively in the experimental and computational literature~\cite{baranov03, dressler2020effect, dressler2020b, dressler2023coexistence, haile26}, with structural derivatives varying protonation state, mixed-anion composition and cross-structure polymorphism, so that the data span deformations and defect environments as well as equilibrium structures.
The second draws on simulation-driven high-throughput screening~\cite{vzguns2024uncovering, haenseroth2026htscreening}, reaching into composition space with little or no experimental precedent: structural motif-based filtering of $6$~million~entries from \textsc{Alexandria}~\cite{alexandria} and the \textsc{Materials Project}~\cite{mp_1, mp_2}, pre-screening with the \textsc{MatterSim} universal MLIP~\cite{mattersim}, AIMD characterization of the survivors and multi-nanosecond simulations with material-specific fine-tuned \textsc{MACE} potentials~\cite{mace_1, mace_2, mace_mp} generated $80{,}000$~DFT calculations per material, while exposing the bottleneck that motivates the present work, namely that a new potential had to be fitted for every candidate~\cite{haenseroth2026htscreening}.
 
\textsc{MACE} was selected as the training framework for its accuracy-to-cost ratio among equivariant graph-network potentials on hydrogen-bonded and molecular systems at moderate training-set sizes~\cite{mace_1, mace_2, mace_mp, kovacs2023}, for its pre-trained universal models~\cite{mace_mp} and well-documented fine-tuning and transfer-learning workflows~\cite{haenseroth2026amaceingtoolkit, radova2025fine, liu2025fine, tompa2026fine}, which lift a subset of the data to range-separated hybrid accuracy without regenerating the full reference set, and for the production MD interfaces and multi-GPU support that make the targeted system sizes and trajectory lengths realistic.
The choice of architecture is nonetheless separable from the contribution of this work: all reference configurations, with their DFT labels at both generalized gradient approximation (GGA) and range-separated hybrid level, are released in a framework-agnostic format and can be used to train or fine-tune potentials of any other architecture, or to extend an existing universal model into the solid acid domain.
Given the pace at which new universal models appear, this curated first-principles data is the component of the present work with the longest expected lifetime~\cite{lahouari2026molcrystmlip}.

Here we present a compound-class-specific MLIP trained exclusively on solid acid materials and assess it in four steps. 
We first quantify force and energy errors against independent DFT calculations that were not part of the training set, and probe generalization along several distinct routes, including zero-shot application to solid acid compositions held out entirely from training. 
We then move beyond force and energy error metrics to physical observables, computing activation energies for proton diffusion in a set of solid acids together with the well-known reorientation dynamics of S- and P-based anions. 
Third, we show that transfer learning extends the model to a higher level of theory with a small fraction of the reference data otherwise required. 
Finally, we introduce a model considerably smaller than previous ones and exploit its reduced cost to run molecular dynamics simulations at system sizes exceeding $250{,}000$~~atoms.
Together these enable the direct computation of converged diffusion coefficients and a mechanistic resolution that has so far remained inaccessible to state-of-the-art universal models~\cite{grunert2025, haenseroth2026amaceingtoolkit, haenseroth2026umlip_config_space}, and they provide a template for local universal model development in other compound classes.

\section*{Results}
\subsection*{Dataset construction}

The dataset comprises $55$~materials in total, which stem from two different sources.
$27$ of these materials were taken from high-throughput materials database screening studies~\cite{vzguns2024uncovering, haenseroth2026htscreening}, in which more than $6$~million~entries of the \textsc{Alexandria}~\cite{alexandria} and \textsc{Materials Project}~\cite{mp_1,mp_2} databases were analyzed by multi-stage workflows combining molecular dynamics simulations with structural motif-based filtering~\cite{vzguns2024uncovering, haenseroth2026htscreening}.
These database entries provide diverse structural configurations across a broad elemental range (see Supplementary Note~1).
The remaining $28$~materials are structural variants of well-known solid acids such as CsH$_2$PO$_4$ and RbH$_2$PO$_4$, which enable a systematic study of proton doping and anion mixing, for example in Cs$_2$(H$_2$PO$_4$)(HSO$_4$)~\cite{dressler2020a, dressler2020effect, dressler2020b, dressler2023coexistence, grunert2025, haile26}.
For each material, $80{,}000$~DFT-labeled configurations were generated from molecular dynamics simulations.
Each simulation was preceded by a geometry optimization of the structure taken from one of the databases or from the literature, construction of the supercell and an equilibration at the material-specific temperature (see Methods).
The average simulation cell in the dataset contains $580$~atoms (see Supplementary Note~2).
The entire dataset therefore comprises $4.4$~million~first-principles calculations.

All oxoanions included in the dataset are shown in Figure~\ref{fig:dataset}a, together with the predominantly alkali and alkaline earth metal cations as well as the positively charged organic cations shown in Figure~\ref{fig:dataset}b.
The dataset spans $19$~elements with atomic numbers up to $81$ (see Supplementary Note~1).
The most frequent cation is Cs$^+$, which occurs in $25$~materials, followed by K$^+$ ($11$) and Rb$^+$ ($7$).
A P-containing anion is present in $45$~\% of the materials, while $31$~\% contain S-, $25$~\% Se- and $13$~\% As-based anionic polyhedra as the oxo-framework; these fractions add up to more than $100$~\% because several compounds contain mixed anions.
All materials are listed in Table~\ref{tab:tab1} together with their database identifiers, where available, and references discussing the respective materials.

\begin{table}[ht]
    \centering
    \caption{
    \textbf{List of solid acid proton conductors included in the dataset.}
    Reported are the composition, the \textsc{Inorganic Crystal Structure Database} (ICSD), \textsc{Materials Project} (MP-ID) or \textsc{Alexandria} (ALEX-ID) identifier, and literature references for the corresponding structures and/or conductivity measurements.
    Multiple entries with the same composition correspond to different polymorphs.
    Dashes indicate structures for which no database entry or reference is available.
    }
    \vspace{3mm}
    {\scriptsize
    \begin{tabular}{lc@{\hspace{1em}}c@{\hspace{1em}}c@{\hspace{1em}}c@{\hspace{1em}}}
        \toprule
        Composition & Database-ID & References \\
        \midrule
        LiH$_{3}$(SeO$_{3}$)$_{2}$ & ALEX-3230254 & \cite{al3230254_struc, al3230254_vib} \\ 
        CsHSeO$_{3}$ & ALEX-3230870 & \cite{al3230870_cond, al3230870_cond2}  \\
        CsHSO$_{4}$ & ALEX-3236698 & \cite{al3236698_struc, al3236698_cond} \\
        Ba(H$_{2}$PO$_{4}$)$_{2}$ & ALEX-3239126 & \cite{gilbert_struct, al3239126_cond} \\
        KH$_{2}$PO$_{4}$ & ALEX-3240111 & \cite{al3240111_struc, al3240111_cond} \\
        KH$_{2}$PO$_{4}$ & ALEX-3240113 & \cite{al3240111_struc, al3240111_cond} \\
        BaHPO$_{4}$ & ALEX-3280448 & \cite{al3280448_struc, al3280448_cond} \\
        Ca(H$_{2}$PO$_{4}$)$_{2}$ & ALEX-5262379 & \cite{al5262379_cond, al5262379_cond2} \\
        RbHSO$_{4}$ & ALEX-5291171 & \cite{al5291171_struc, al5291171_cond} \\
        K$_{2}$H$_{2}$As$_{2}$O$_{7}$ & ALEX-6308766 & -- \\
        (KRb$_{3}$)(H$_{2}$As$_{2}$O$_{7}$)$_2$ & ALEX-6308815 & -- \\
        (KCs$_{3}$)(H$_{2}$P$_{2}$O$_{7}$)$_2$ & ALEX-6308862 & -- \\
        (KCs$_{3}$)(H$_{2}$As$_{2}$O$_{7}$)$_2$ & ALEX-6308874 & -- \\
        Rb$_{4}$H$_{4}$(Se$_{3}$S)O$_{16}$ & ALEX-6308968 & -- \\
        RbHSeO$_{4}$ & ALEX-6308988 & \cite{al6308988_struc, al6308988_cond} \\
        Sn(H$_{2}$PO$_{4}$)$_{2}$ & ALEX-7541875 & \cite{al7541875_struc} \\
        CsHSeO$_{3}$ & ALEX-7647189 & \cite{al3230870_cond, al3230870_cond2} \\
        KH$_{3}$(SeO$_{3}$)$_{2}$ & ALEX-7754128 & \cite{al7754128_struc_cond} \\
        
        Rb$_2$H$_6$(SO$_4$)(TeO$_6$) & MP-559096 & \cite{mp559096_struc_cond} \\
        KH$_{2}$PO$_{4}$ & MP-699437 & \cite{al3240111_struc} \\
        Tl$_2$H$_6$(SeO$_4$)(TeO$_6$) & MP-1196566 & \cite{mp1196566_struc_cond} \\
        K$_{4}$LiH$_{3}$(SO$_{4}$)$_{4}$ & MP-1196827 & \cite{mp1196827_cond, mp1196827_cond2} \\
        H$_{56}$C$_{19}$S$_3$N$_8$O$_{15}$ & MP-1196861 & \cite{mp1196861_struc} \\
        KHSO$_{4}$ & MP-1199797 & \cite{mp1199797_cond, mp1199797_cond2} \\
        (CH$_{3}$)$_{2}$NH$_{2}$H$_{2}$PO$_{4}$ & MP-1200282 & \cite{mp1200282_struc, mp1200282_cond}  \\
        (NH$_{4}$)$_{3}$(HSeO$_{4}$)$_{2}$ & MP-1212963 & -- \\
        (Al(H$_2$O)$_6$)$_4$H$_4$(SO$_4$)$_7$ & MP-1229278 & -- \\
        
        CsH$_2$PO$_4$                  & ICSD-151916 & \cite{struct_cdp} \\
        Cs$_{63}$H(H$_2$PO$_4$)$_{64}$                & -- & \cite{struct_cdp} \\
        Cs$_{31}$H(H$_2$PO$_4$)$_{32}$                & -- & \cite{struct_cdp}  \\
        Cs$_{15}$H(H$_2$PO$_4$)$_{16}$                & -- & \cite{struct_cdp}  \\
        Cs$_{7}$H(H$_2$PO$_4$)$_{8}$                & -- &  \cite{struct_cdp}  \\
        Cs$_7$(H$_4$PO$_4$)(H$_2$PO$_4$)$_8$  & -- & \cite{struct_cpp} \\
        CsHSO$_4$                      & ICSD-63352 & \cite{chs_struct}\\
        CsHSO$_4$ (cubic sym.)                & -- & \cite{struct_cdp,chs_struct} \\
        Cs$_2$(H$_2$PO$_4$)(HSO$_4$) (conf. i)                 & -- & \cite{baranov03,struct_cdp,chs_struct} \\
        Cs$_2$(H$_2$PO$_4$)(HSO$_4$) (conf. ii)                 & -- & \cite{baranov03,struct_cdp,chs_struct} \\
        Cs$_2$(H$_2$PO$_4$)(HSO$_4$) (conf. iii)                 & -- & \cite{baranov03,struct_cdp,chs_struct} \\
        Cs$_3$(HSO$_4$)$_2$(H$_2$PO$_4$) & ICSD-79789 & \cite{Haile:du0399} \\
        Cs$_6$(SO$_4$)$_3$(H$_3$PO$_4$)$_4$               & -- & \cite{Makarova:yh5010} \\
        Cs$_4$(HSO$_4$)$_3$(H$_2$PO$_4$)         & ICSD-128940 & \cite{Makarova:lo5120} \\
        CsHSeO$_4$ & ICSD-411281 & \cite{chse_struc} \\
        Cs$_3$H(SeO$_4$)$_2$               & -- & \cite{merinov_bolotina_baranova_shuvalov_1991} \\
        Cs$_2$(H$_2$PO$_4$)(HSeO$_4$)                  & -- & \cite{struct_cdp,baranov03,chse_struc} \\
        CsH$_2$AsO$_4$                  & ICSD-201167 & \cite{WJHay_1981} \\
        Cs(H$_2$AsO$_4$)(H$_3$AsO$_4$)$_2$  & ICSD-141985 & \cite{Schwendtner:qf3024} \\
        Cs$_4$(SeO$_4$)(HSeO$_4$)$_2$(H$_3$AsO$_4$)               & -- & \cite{AMRI200968} \\
        (NH$_4$)$_3$H(SeO$_4$)$_2$               & -- & \cite{Lukaszewicz:na0025} \\
        NH$_4$(H$_2$AsO$_4$)(H$_3$AsO$_4$)       & ICSD-141983 & \cite{Schwendtner:qf3024} \\
        K$_4$(NH$_4$)$_5$H$_3$(SO$_4$)$_6$               & -- & \cite{Selezneva:yh5025} \\ 
        Li$_2$(H$_2$PO$_4$)$_2$               & ICSD-141984 & \cite{Schwendtner:qf3024} \\
        Rb$_3$H(SeO$_4$)$_2$               & -- & \cite{Magome15042009} \\
        RbH$_2$PO$_4$               & -- & \cite{Botez_2009} \\
        ZrH$_5$(PO$_4$)$_3$               & -- & \cite{zrh5po43} \\
        SnHPO$_4$               & -- & \cite{snhpo34_mat} \\
    \bottomrule
    \end{tabular}
    }
    \label{tab:tab1}
\end{table}

\subsection*{Model training, uncertainty and validation}

Two domain-specific machine learning interatomic potentials were prepared following the design principles established for the \textsc{MACE-MP-0} models.
The larger of the two, named `medium' in line with the naming convention of the other \textsc{MACE} universal models, incorporates equivariant messages in the message-passing algorithm ($\max L = 1$), whereas the smaller one, named `small', uses invariant messages only ($\max L = 0$).
The parameter $\max L$ determines which type of messages are passed through the network (see Methods).
A hyper-parameter optimization was carried out prior to training (see Supplementary Note~3), and the resulting hyper-parameters are listed in Table~\ref{tab:tab2}.

Because the number of known solid acid materials is small, a large number of configurations per material was chosen to obtain accurate models.
The solid acid dataset was sub-sampled equidistantly.
For the `medium' model a stride of $5$ was applied, resulting in a training set of $880{,}000$~configurations ($16{,}000$~per material).
The corresponding validation set was constructed by randomly sub-sampling $3{,}000$ of the remaining $64{,}000$~DFT configurations per material ($165{,}000$ configurations in total).
The test set used throughout the following evaluations consists of $200$ configurations drawn at random from the remaining $61{,}000$~configurations per material ($11{,}000$ configurations in total).
The training and validation sets for the `small' model were obtained by taking every second configuration of the corresponding `medium' sets, yielding $440{,}000$ and $82{,}500$~configurations, respectively ($8{,}000$ and $1{,}500$~per material).
The same test set was used for the evaluation of the `small' model.

To quantify the uncertainty of the reported errors, a committee of five `small' and three `medium' models with different seeds was trained~\cite{Wen2020, lu2023uncertaintyestimatormlip, Janet2019, Kellner_2024}; all error values below are given as the committee mean and standard deviation.
The force and energy mean absolute errors (MAEs) of the resulting models were evaluated on the test set of each material (see Supplementary Notes~5 and 6).
For comparison, the corresponding errors of the universal models \textsc{MACE-MP-0} (small)~\cite{mace_mp}, \textsc{MACE-OMAT-0} (medium)~\cite{batatia2025cross} and \textsc{PET-OAM-XL}~\cite{pet_oam} are reported in Supplementary Notes~7, 8 and~9, respectively.
Figure~\ref{fig:fig2}a provides an overview of the force errors of the universal models and of the domain-specific \textsc{SolidAcid-0} `small' and `medium' models, alongside a material-specific \textsc{MACE} model trained from scratch (on the `small' training set of $8{,}000$~configurations), evaluated on a subset of $10$~randomly drawn materials.
The material-averaged errors of the \textsc{SolidAcid-0} `medium' model are $0.01109 \pm 0.00001$~eV/{\AA} for the forces and $0.00214 \pm 0.00005$~eV/atom for the energies.
The best-performing universal model on the test sets is \textsc{PET-OAM-XL}, followed by \textsc{MACE-OMAT-0} (medium) and \textsc{MACE-MP-0} (small), the best reaching a force error of $0.13095$~eV/{\AA}.
The material-specific models perform similarly to, or slightly better than, the \textsc{SolidAcid-0} `medium' model, at about $0.01$~eV/{\AA}.

The \textsc{SolidAcid-0} models are accurate across the full set of materials, with reduced performance only for the more artificial structures such as `CsHSO$_4$ (cubic)', in which the sulfate compound was constructed artificially within the cubic cell of CsH$_2$PO$_4$ (see Supplementary Notes~5 and~6).
Elevated energy MAEs are likewise found for the structural derivatives of CsH$_2$PO$_4$ in which every $n$-th Cs$^+$ is artificially replaced in the simulation by an H$^+$ ($n\in\{8,16,32,64\}$).
The improvement of the `medium' over the `small' model is of comparable magnitude for all materials, amounting on average to a $23$~\% lower force error and a $3$~\% lower energy error.

To probe the out-of-domain capabilities of the \textsc{SolidAcid-0} MLIPs, three additional `small' models were trained with the training and validation configurations of a randomly drawn subset of $12$~materials removed, using a different subset in each case.
These models were thus trained on only $78$~\% of the original `small' dataset ($344{,}000$~configurations).
Their performance on the individual materials is reported in Supplementary Note~10.
On the held-out compounds the models are $56$~\% worse in the median and $120$~\% worse on average, corresponding to a mean force error of $0.0319$~eV/{\AA}, while their errors on the retained materials remain comparable to those of the `small' model trained on the full dataset.
This behavior points to a local generalization of the investigated solid acid MLIPs: predictive accuracy decreases for unseen compounds, but for structures closely related to those in the training set the models still outperform state-of-the-art universal models.
Within this restricted chemical neighborhood, forces and energies for unseen materials are therefore predicted with an accuracy approaching that obtained for the materials included in training.
These results allow a cautious estimate of how the models can be expected to behave on solid acids that are not part of the training set.

\subsection*{Proton diffusion and anion rotation dynamics}

In addition to the force~\cite{fu2022forces} and energy MAEs, molecular dynamics simulations were performed with the resulting models for twelve randomly selected materials.
The radial distribution functions (RDFs) of the O--O and O--H pairs obtained with the \textsc{SolidAcid-0} `small' and `medium' models are compared with those of \textsc{MACE-MP-0} (small) and with the first-principles reference in Supplementary Note~11.
While the universal model already provides a good description in many cases, the \textsc{SolidAcid-0} models reproduce the AIMD reference closely for all compounds considered.

\begin{figure*}[ht]
    \centering
    \includegraphics{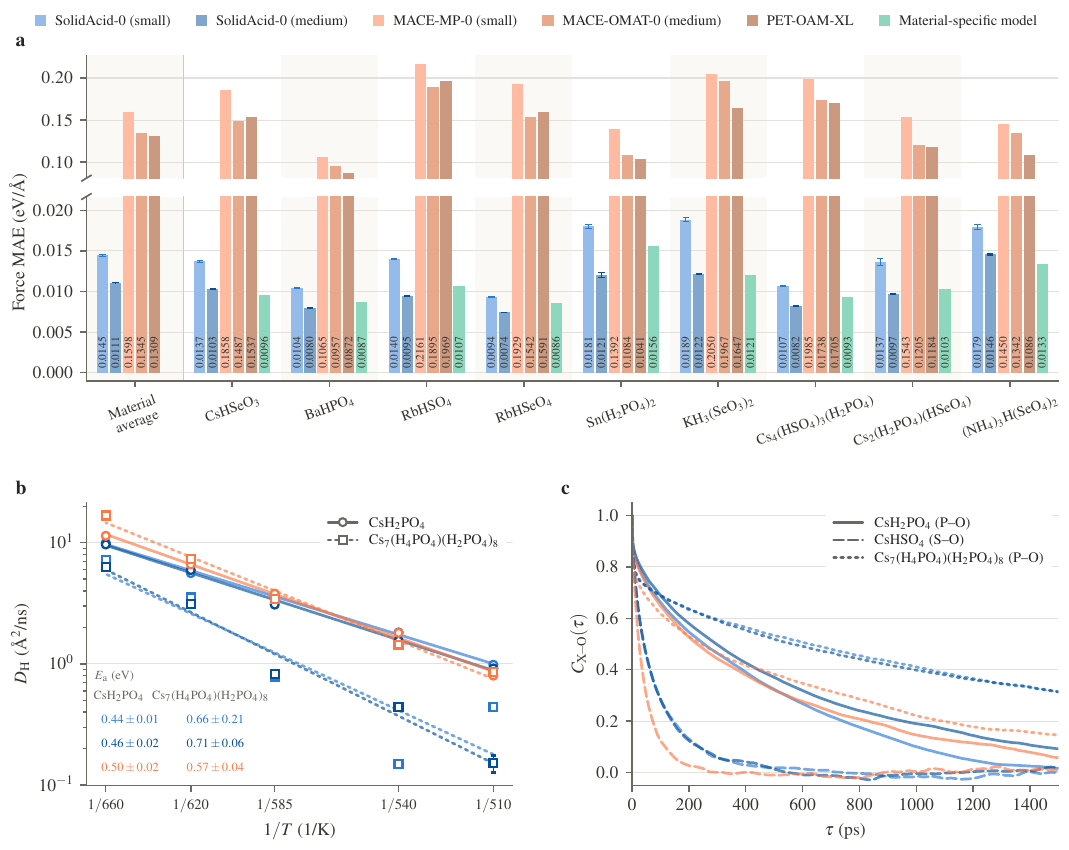}
    \caption{
    \textbf{Accuracy and proton transport properties of the \textsc{SolidAcid-0} models.}
    \textbf{a} Force mean absolute errors on the test sets of nine randomly drawn materials and material averaged errors (all $55$~materials) for the universal models \textsc{MACE-MP-0} (small), \textsc{MACE-OMAT-0} (medium) and \textsc{PET-OAM-XL}, the domain-specific \textsc{SolidAcid-0} `small' and `medium' models, and a \textsc{MACE} model trained from scratch on the single material ($8{,}000$~configurations).
    Errors for the full set of $55$~materials are reported in Supplementary Notes~4 to~9.
    Error bars denote the standard deviation over the model committee; for the material-averaged bars the average over the $55$~materials is formed per committee member before the standard deviation is taken, i.e. it is the committee spread, not the spread between materials.
    \textbf{b} Arrhenius plots of the proton diffusion coefficients of CDP (solid lines) and CPP (dashed lines) obtained from the $3$~ns trajectories, with the resulting activation energies obtained with \textsc{MACE-OMAT-0} (medium), \textsc{SolidAcid-0} `small' and `medium'.
    \textbf{c} X--O vector autocorrelation functions describing the anion reorientation dynamics in CDP, CPP and CHS at $510$~K obtained with \textsc{MACE-OMAT-0} (medium), \textsc{SolidAcid-0} `small' and `medium'.
    The underlying mean square displacements are shown in Supplementary Notes~12 to~15.
    }
    \label{fig:fig2}
\end{figure*}

For the well-known solid acids CsH$_2$PO$_4$ (CDP), CsHSO$_4$ (CHS) and Cs$_7$(H$_4$PO$_4$)(H$_2$PO$_4$)$_8$ (CPP), molecular dynamics trajectories of several nanoseconds were generated with the two \textsc{SolidAcid-0} models as well as with \textsc{MACE-MP-0} (small) and \textsc{MACE-OMAT-0} (medium).
For CDP and CPP the activation energies of the proton transport process were determined (Figure~\ref{fig:fig2}b).
The experimental reference values are $0.41$~eV for CDP~\cite{baranov03} and $0.65$~eV for CPP~\cite{struct_cpp}.
The \textsc{SolidAcid-0} models reproduce these experimental values more accurately than the universal \textsc{MACE} models.

Besides proton diffusion, the anion reorientation dynamics were investigated for the same three compounds (Figure~\ref{fig:fig2}c).
The S--O vectors in CHS are known to reorient faster than the P--O vectors in CDP, and the reorientation in CPP is slower still~\cite{grunert2025, dressler2020effect, dressler2023coexistence}.
This ordering is reproduced by the \textsc{SolidAcid-0} models, whereas the universal \textsc{MACE-MP-0} (small) model predicts similarly fast reorientation dynamics for the P--O vectors in CDP and CPP.

\subsection*{Transfer learning}

A second dataset was generated at the hybrid-functional DFT level using the range-separated HSE06 functional~\cite{hse06_1, hse06_2}, comprising $550$~configurations for each of CsH$_2$PO$_4$, CsHSO$_4$, CsH$_2$AsO$_4$ and CsHSeO$_4$.
The \textsc{SolidAcid-0} `small' model was transfer-learned to this level of theory using the `naive' fine-tuning approach~\cite{tompa2026fine,haenseroth2026amaceingtoolkit}, separately for each target material.
The force and energy errors of the transfer-learned \textsc{SolidAcid-0} `small' models are reported in Supplementary Note~17, and their predictions for the hybrid-functional O--H and O--O RDFs in Supplementary Note~18.
The material-averaged force MAE is $0.0244\pm0.0006$~eV/{\AA} and the energy MAE is $0.0005\pm0.0001$~eV/atom (see Supplementary Note~17).
The radial distribution functions of the four compounds are recovered by the transfer-learned \textsc{SolidAcid-0} models (see Supplementary Note~18).
The transfer-learned models were further used to run molecular dynamics simulations of several nanoseconds, from which the anion reorientation dynamics and the activation energies shown in Figure~\ref{fig:fig3}c were extracted.
The activation energy obtained for CDP is $0.43\pm0.02$~eV, which compares to the experimental value of $0.41$~eV~\cite{baranov03} (see Supplementary Note~19).

\subsection*{Increased simulation speed and enlarged simulation cells}

\begin{figure*}[ht]
    \centering
    \includegraphics{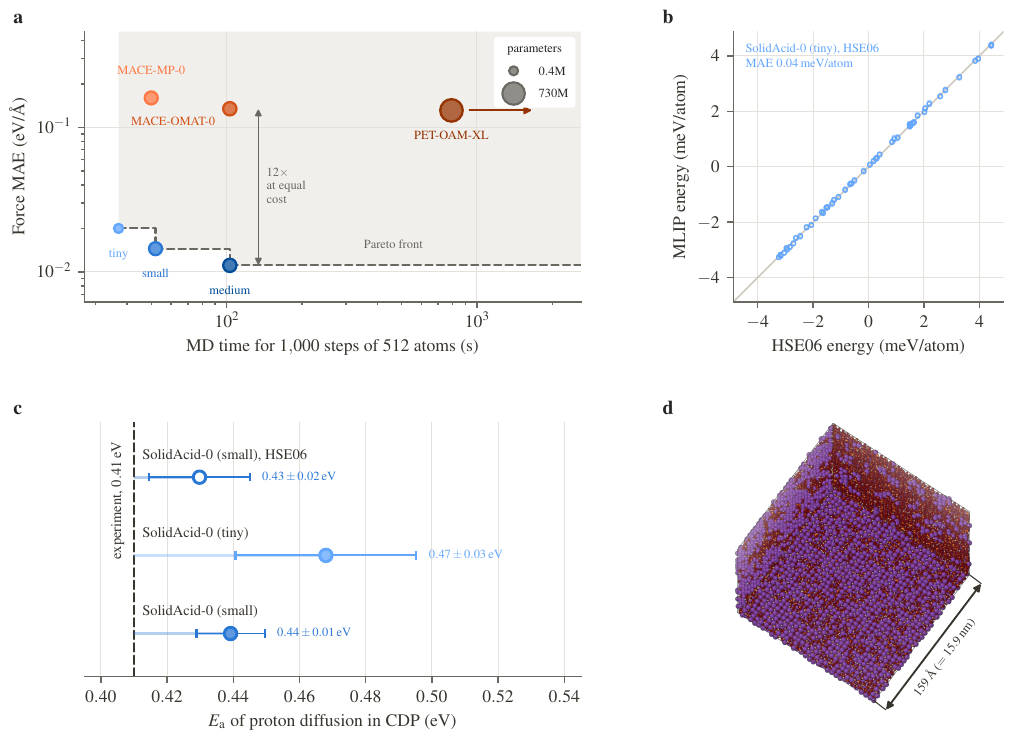}
    \caption{
    \textbf{A smaller model, transferred to hybrid DFT, and the cell size it opens up.}
    \textbf{a} Speed against accuracy of the \textsc{SolidAcid-0} and universal models: molecular-dynamics time for $1{,}000$ steps of a $512$-atom cell at \texttt{float64} on one \textsc{NVIDIA} A100 (PCIe, $250$~W, $40$~GB), against the force mean absolute error averaged over the $55$-material test set; the marker area scales with the number of parameters. 
    Among the models benchmarked here, the three \textsc{SolidAcid-0} models are the entire Pareto front (dashed), and every universal model lies inside the shaded region above and to the right of it. 
    At the same cost of $103$~s, \textsc{SolidAcid-0} (medium) is $12$ times more accurate than
    \textsc{MACE-OMAT-0}, with which it shares its architecture. 
    \textsc{PET-OAM-XL} was benchmarked only at \texttt{float32}, so a corresponding \texttt{float64} time would lie further to the right (arrow). 
    Error bars are the spread of the material-averaged error over five training seeds; the universal models are single-seed evaluations.
    \textbf{b} Predicted against reference energy per atom for the \textsc{SolidAcid-0} (tiny) model transfer-learned to HSE06, over the $50$ configurations of the CsH$_2$PO$_4$ test set and all five seeds. 
    Both axes are referenced to the mean of the test set.
    \textbf{c} Activation energy of proton diffusion in CDP, drawn as the distance from the measured value of $0.41$~eV~\cite{baranov03}. 
    Transfer learning to HSE06 brings the prediction onto the experiment. 
    Unmarked models are at the PBE level; the hollow marker is the transfer-learned one. 
    Uncertainties come from the $\ln D$ against $1/T$ regression.
    \textbf{d} Last frame of a $262{,}144$-atom CsH$_2$PO$_4$ simulation at $510$~K run with the transfer-learned \textsc{SolidAcid-0} (tiny) model on a single \textsc{NVIDIA} H200 ($141$~GB), an $32\!\times\!32\!\times\!32$ replication of the cubic unit cell of CDP. 
    Cesium purple, oxygen red, phosphorus orange, hydrogen grey.   
    }
    \label{fig:fig3}
\end{figure*}

To extend the limits of current MLIP-based molecular dynamics simulations, a third model, \textsc{SolidAcid-0} `tiny', was prepared.
It shares the hyper-parameters of the \textsc{SolidAcid-0} `small' model and of the other \textsc{MACE} `small' models, but uses $64\!\times\!0e$ instead of $128\!\times\!0e$ hidden irreps, which reduces the number of parameters by $78$~\% ($432{,}276$ instead of $1{,}976{,}864$).
With this model a speed-up of up to $30$~\% relative to the `small' model was achieved in MD simulations (see Fig.~\ref{fig:fig3}a and Supplementary Note~23).
Its material-averaged error is higher than that of the two larger \textsc{SolidAcid-0} models, but still lower than that of the state-of-the-art universal models: material-averaged force MAE is $0.0200\pm0.0003$~eV/{\AA} and the energy MAE is $0.0024\pm0.0001$~eV/atom (see Supplementary Note~20).
The radial distribution functions of the twelve randomly selected materials are recovered by the \textsc{SolidAcid-0} (tiny) model with similar accuracy as the \textsc{SolidAcid-0} (small) model (see Supplementary Note~21).
After transfer learning the `tiny' model to the HSE06 level for CDP, MD simulations of CDP with more than $250{,}000$~atoms became feasible on a single \textsc{NVIDIA} H200 GPU ($141$~GB; see Figure~\ref{fig:fig3}b and d).
Such cells correspond to $32$~unit cells of CDP and a box length of $159$~{\AA}.
Simulations at this scale can provide mechanistic insight into proton transport and collective dynamics at length scales approaching experimental conditions.

\section*{Discussion}

Universal MLIPs are trained for breadth, and breadth is not the property that solid state proton transport requires.
Conduction in these materials rests on a rare-event sequence, a Grotthuss transfer between neighboring oxoanions followed by the anion reorientation that sets up the next one, which a chemically heterogeneous training distribution samples only thinly.
We therefore constructed the training distribution around that process: $4.4$~million~DFT-labeled configurations spanning $55$~solid acid materials, drawn from AIMD trajectories of each compound.
The resulting \textsc{SolidAcid-0} models outperform \textsc{PET-OAM-XL}, \textsc{MACE-OMAT-0} (medium) and \textsc{MACE-MP-0} (small) throughout this domain~\cite{pet_oam, batatia2025cross, mace_mp}, reaching a material-averaged force error of $0.011$~eV/{\AA} with a fraction of the parameters of the largest of them.

The clearest evidence that domain specificity matters here is that static and dynamic accuracy come apart.
The universal models already reproduce the O--O and O--H radial distribution functions of most compounds reasonably well, yet \textsc{MACE-MP-0} (small) predicts near-identical P--O reorientation dynamics for CDP and CPP, an ordering that earlier simulations contradict~\cite{dressler2020effect, dressler2023coexistence, grunert2025}.
A potential can therefore be accurate on structure and still be wrong about the observable of interest, which supports the broader argument that force and energy MAEs alone are an incomplete criterion for model quality~\cite{fu2022forces}.
The price of covering $55$~compositions within a single model is meanwhile small: \textsc{MACE} models trained from scratch on an individual material are only marginally better than \textsc{SolidAcid-0} `medium', so multi-material training buys transferability at almost no cost in per-material accuracy.

The limits of that transferability should be stated plainly.
On compositions removed entirely from training, the models degrade to a force error of $0.0319$~eV/{\AA}, which is not the behavior of a universal potential.
What remains is a local generalization: for compounds structurally close to the training set, zero-shot accuracy stays competitive with state-of-the-art universal models, while the errors on the retained materials are essentially unaffected.
In practice this positions \textsc{SolidAcid-0} as a screening tool within the neighborhood of known solid acids, covering new A-site substitutions, mixed-anion compositions and protonation variants, rather than as a general-purpose potential for arbitrary liquid and solid state hydrogen-bonded chemistry.
Extending its reach is consequently a question of sampling a wider range of local environments, not of changing the architecture~\cite{elmachachi2024gomace, mahmoud2025mace_go}.

Two further results decouple quantities that have historically traded against each other.
Fine-tuning to the HSE06 level required only $500$~configurations per material, so hybrid-functional accuracy no longer demands a large hybrid-functional dataset; and the `tiny' model, with $78$~\% fewer parameters and up to $30$~\% faster inference, sustains cells of more than $260{,}000$~atoms on a single GPU.
The level of theory of the reference and the length scale of the simulation can thus be selected largely independently of one another.
This brings within reach phenomena that AIMD cells could address: domain walls, grain boundaries and the superprotonic transition itself, all of which involve correlated motion over length scales far beyond a few hundred atoms.

The dataset covers bulk crystalline phases only, without surfaces, interfaces or realistic defect populations, and the majority of it is labeled at the PBE level; the AIMD sampling spans $400$ to $510$~K.
All MLIP MD simulations treat the nuclei classically, which is a genuine approximation for a mechanism built on the motion of a single proton, as zero-point energy and tunneling are known to lower transfer barriers in hydrogen-bonded solids~\cite{marx2000ab, tuckerman2002ab, marx1996, habershon2007, ceriotti2016}.
Path-integral molecular dynamics would capture these nuclear quantum effects but multiplies the cost by the number of beads, and this is precisely where a cheap yet domain-accurate model pays off: the \textsc{SolidAcid-0} `tiny' model makes bead counts tractable that would be prohibitive with larger potentials, while remaining more accurate on this chemistry than any universal model tested here.
The dataset and all three models are deposited openly, as a starting point for active-learning extensions into new solid acid chemistries, for fine-tuning to higher levels of theory, and for training architectures other than \textsc{MACE}.

\section*{Methods}
 
\subsection*{Ab initio molecular dynamics simulations}
 
The dataset generation relied on ab initio molecular dynamics simulations to obtain DFT-labeled configurations.
The reference calculations for the \textsc{SolidAcid-0} models were performed with \textsc{CP2K} (version~2025.1)~\cite{cp2k_1, cp2k_2, cp2k_3, cp2k_4, cp2k_5, cp2k_quickstep, cp2k_orb_trans}, using the PBE exchange-correlation functional~\cite{pbe}, GTH pseudo-potentials~\cite{cp2k_gth-pseudopot1, cp2k_gth-pseudopot2, cp2k_gth-pseudopot3}, the DZVP-MOLOPT basis set~\cite{cp2k_basis-set}, and a Nos\'{e}-Hoover chain thermostat~\cite{nose1, nose2, nose3}.
For each material, the production simulation from which the dataset configurations were extracted was preceded by a geometry optimization and a $2{,}000$-step equilibration.
The AIMD simulations were carried out for $40$~ps at material-specific temperatures ranging from $400$ to $510$~K, with a timestep of $0.5$~fs.

The higher-level-of-theory dataset was generated with the range-separated hybrid functional HSE06~\cite{hse06_1, hse06_2}.
For CsH$_2$PO$_4$, CsHSO$_4$, CsH$_2$AsO$_4$ and CsHSeO$_4$ molecular dynamics simulations were conducted with the same calculation parameters as the GGA-level AIMD, but smaller simulation cells, yielding $550$ DFT-labeled configurations per material.

\subsection*{MLIP training and molecular dynamics simulations}
 
The models were trained with the \textsc{MACE}-torch package (version~0.3.14)~\cite{mace_1, mace_2}.
Hyper-parameter optimization was performed for the `small' model, and the resulting hyper-parameters were used for all model sizes (see Table~\ref{tab:tab2} and Supplementary Note~3).
For each element present in the PBE and HSE06 dataset, isolated single-atom \textsc{CP2K} calculations were performed with the same exchange-correlation functional, basis sets, pseudo-potentials, and numerical settings as used for the reference trajectories. 
Spin-polarized calculations were used.
The resulting $E_0$ values were supplied as atomic energy offsets during training and transfer-learning and are reported in the Table~\ref{tab:tab2} and Supplementary Note~16.
Training was conducted on \textsc{NVIDIA} H200 GPUs for the three `medium' models and on \textsc{NVIDIA} A100 GPUs for all other model sizes.
 
The models were evaluated in terms of physical observables and of the mean absolute errors of the energies and forces, defined as
 
\begin{equation*}
	\operatorname{MAE}(E) = \frac{1}{N M} \sum_{s=1}^{N} \left\vert E_s^{\mathrm{pred}} - E_s^{\mathrm{ref}} \right\vert ,
\end{equation*}

\begin{equation*}
        \operatorname{MAE}(F) = \frac{1}{3 N M} \sum_{s=1}^{N} \sum_{a=1}^{M}  \sum_{\alpha \in \{x,y,z\}} \vert F_{s,a,\alpha}^{\mathrm{pred}} -
 F_{s,a,\alpha}^{\mathrm{ref}} \vert ,
\end{equation*}
 
where $N$ is the number of configurations, $M$ the number of atoms per configuration and $\alpha$ the Cartesian components, giving errors in eV/atom and eV/{\AA}, respectively.

The errors are reported separately for each material, together with the material-averaged values, both for the models presented here and for the universal models \textsc{MACE-MP-0} (small), \textsc{MACE-OMAT-0} (medium) and \textsc{PET-OAM-XL}, in Supplementary Notes~7, 8 and~9.

\begin{table}[h!]
    \centering
    \caption{
    \textbf{Model overview.}
    hyper-parameters, reference energies and dataset sizes of the three domain-specific \textsc{SolidAcid-0} \textsc{MACE} models.
    }
    \vspace{3mm}
    \begin{tabular}{lc@{\hspace{1em}}c@{\hspace{1em}}c@{\hspace{1em}}}
        \toprule
        Model size & tiny & small & medium \\
        \midrule
	    \multicolumn{4}{l}{\textit{Hyper-parameter}} \\
        Learning rate & $0.005$ & $0.005$ & $0.005$ \\
        Weight decay & $10^{-8}$ & $10^{-8}$ & $10^{-8}$ \\
        EMA decay & $0.999$ & $0.999$ & $0.999$ \\
        Batch size & $8$& $8$ & $8$ \\
        Force-energy-loss ratio & $10$ & $10$ & $10$ \\
        Number of epochs & $25$ & $25$ & $25$ \\
        $\max L$ & $0$ &  $0$ & $1$ \\
        Messages & $64 \times 0e $ & $128 \times 0e $ & $128 \times 0e$ \\ 
        &&& $+128 \times 1o $ \\ 
        Message passing layers & $2$ & $2$ & $2$ \\
        Cutoff radius (\AA) & $6.0$ & $6.0$ & $6.0$ \\
        Angular resolution ($\ell$) & $2$ & $3$ & $3$ \\
        Readout MLP irreps & $8\times0e$ & $16\times0e$ & $16\times0e$ \\
        Radial basis type & Bessel & Bessel & Bessel \\
        No. of radial basis fct. & 6 & 8 & 8 \\
        Polyn. cutoff order & 4 & 5 & 5 \\
        Float precision & \texttt{float64} & \texttt{float64} & \texttt{float64} \\
        \textsc{MACE} model & \texttt{MACE} & \texttt{MACE} & \texttt{MACE} \\
        1st interaction type & RADIB & RADIB & RADIB \\
        2nd interaction type & RADRIB & RADRIB & RADRIB \\
        Correlation order & $3$ & $3$ & $3$ \\
        Optimizer~\cite{kingma2014adam} & Adam & Adam & Adam \\
        Seeds & $1111$ & $1111$ & $1111$ \\
        & $2222$ & $2222$ & $2222$ \\
        & $3333$ & $3333$ & $3333$ \\
        & $4444$ & $4444$ & -- \\
        & $5555$ & $5555$ & -- \\
	\midrule
	\multicolumn{4}{l}{\textit{Dataset size}} \\
	Training set size & $440,000$ & $440,000$ & $880,000$ \\
	Validation set size & $82{,}500$ & $82{,}500$ & $165{,}000$ \\
	Test set size & $11{,}000$ & $11{,}000$ & $11{,}000$ \\
	\midrule
        \multicolumn{4}{l}{\textit{Reference energies} (eV)} \\
        \hspace{3em} H & \multicolumn{3}{c}{-12.54137} \\
        \hspace{3em} Li & \multicolumn{3}{c}{-200.44969} \\
        \hspace{3em} C & \multicolumn{3}{c}{-145.69842} \\
        \hspace{3em} N & \multicolumn{3}{c}{-264.34984} \\
        \hspace{3em} O & \multicolumn{3}{c}{-429.98006} \\
        \hspace{3em} Al & \multicolumn{3}{c}{-52.85190} \\
        \hspace{3em} P & \multicolumn{3}{c}{-175.14050} \\
        \hspace{3em} S & \multicolumn{3}{c}{-272.63597} \\
        \hspace{3em} K & \multicolumn{3}{c}{-767.43449} \\
        \hspace{3em} Ca & \multicolumn{3}{c}{-996.46692} \\
        \hspace{3em} As & \multicolumn{3}{c}{-168.01855} \\
        \hspace{3em} Se & \multicolumn{3}{c}{-252.15096} \\
        \hspace{3em} Rb & \multicolumn{3}{c}{-654.65188} \\
        \hspace{3em} Zr & \multicolumn{3}{c}{-1274.84176} \\
        \hspace{3em} Sn & \multicolumn{3}{c}{-91.63989} \\
        \hspace{3em} Te & \multicolumn{3}{c}{-310.64083} \\
        \hspace{3em} Cs & \multicolumn{3}{c}{-546.98069} \\
        \hspace{3em} Ba & \multicolumn{3}{c}{-691.08459} \\
        \hspace{3em} Tl & \multicolumn{3}{c}{-1355.11243} \\
    \bottomrule
    \multicolumn{4}{l}{{\scriptsize RADIB: \texttt{RealAgnosticDensityInteractionBlock}}}\\
    \multicolumn{4}{l}{{\scriptsize RADRIB: \texttt{RealAgnosticDensityResidualInteractionBlock}}}\\
    \end{tabular}
    \label{tab:tab2}
\end{table}

Molecular dynamics simulations were performed with \textsc{ASE} (version 3.25), employing Nos\'{e}-Hoover chain thermostats set up with the \textsc{aMACEing\_toolkit}~\cite{mace_1, mace_2, ase, nose1, nose2, nose3, haenseroth2026amaceingtoolkit}.
Inference was carried out in \texttt{float32} precision using the model trained with seed 1111 for the production MD simulations.
The MD simulations were run for $3$~ns, for CDP and CPP at $510$, $540$, $585$, $620$ and $660$~K and for CHS at $510$~K, using a timestep of $0.5$~fs, a thermostat chain length of $3$ and a damping factor of $50$~fs ($100$~timesteps).
The calculations were executed on the GPU cluster of Technische Universit{\"a}t Ilmenau using \textsc{NVIDIA} A100 GPUs; the large simulation cells were run on \textsc{NVIDIA} H200 GPUs.

\subsection*{Trajectory analyses}

Converged diffusion coefficients were obtained with the \textsc{SolidAcid-0} models, following the protocols previously established for the solid acids CsH$_2$PO$_4$ and Cs$_7$(H$_4$PO$_4$)(H$_2$PO$_4$)$_8$~\cite{grunert2025}.
 
The dynamical behavior of the systems was analyzed by calculating the mean square displacement (MSD) of the protons in the trajectories of both the \textsc{SolidAcid-0} and the universal \textsc{MACE} models, according to Equation~(\ref{eq:msd}).
Here, $\underline{r}_{i}(t)$ denotes the position of particle $i$ at time $t$.
The $\mathrm{MSD}(\tau)$ was computed by averaging over all particles $i$ and all initial times $t$ satisfying $t + \tau < t_\mathrm{MD}$, where $t_\mathrm{MD}$ is the total simulation time.
\begin{equation}
    \mathrm{MSD}(\tau) = \langle \left\vert \underline{r}_{i}(t+\tau) - \underline{r}_{i}(t) \right\vert^{2} \rangle_{t,i}
    \label{eq:msd}
\end{equation}
Diffusion coefficients $D$ were derived from the MSD data as shown in Eq.~(\ref{eq:diff_coeff}), by linear fitting within an appropriate time interval $\tau \in [100~\mathrm{ps},\,1{,}000~\mathrm{ps}]$ (see Supplementary Notes~12-15).
\begin{equation}
    D = \frac{1}{6} \frac{d}{d \tau}\mathrm{MSD}(\tau)
    \label{eq:diff_coeff}
\end{equation}
The exceptionally high proton diffusivity observed in these solid compounds is connected to their ability to sustain structural diffusion via the Grotthuss mechanism.
Proton transport proceeds through transfer events between two XO$_z$ groups: a proton that is covalently bound to one group and hydrogen-bonded to the other exchanges these two bonding roles and thereby hops between the groups.
The number of proton transfer events per picosecond and per hydrogen atom was evaluated.
The second step of the Grotthuss mechanism in solid acids is the reorientation of the XO$_z$ groups, which enables subsequent proton transfer events with different XO$_z$ partners.
The rotational dynamics of the XO$_z$ groups were analyzed using the X--O vector autocorrelation function $C_{\mathrm{X-O}}(\tau)$, defined in Eq.~(\ref{eq:vacf}).
Here, $\hat{\underline{r}}_{\mathrm{X-O}}$ is the unit vector along the X--O bond.
\begin{equation}
    C_{\mathrm{X-O}}(\tau) = \langle \langle \hat{\underline{r}}_{\mathrm{X-O}}(t_0) \cdot \hat{\underline{r}}_{\mathrm{X-O}}(t_0 + \tau) \rangle_{\mathrm{XO}} \rangle_{t_0}
    \label{eq:vacf}
\end{equation}

Activation energies for proton diffusion were extracted from the temperature dependence of the diffusion coefficients calculated at $510$, $540$, $585$, $620$ and $660$~K from the $3$~ns molecular dynamics trajectories.
The relation between $D$ and $T$ follows an Arrhenius-type behavior as defined in Eq.~(\ref{eq:act_ener}), with prefactor $A$, activation energy $E_a$, Boltzmann constant $k_\mathrm{B}$ and temperature $T$.
\begin{equation}
    D(T) = A \exp\left( -\frac{E_a}{k_\mathrm{B} T} \right)
    \label{eq:act_ener}
\end{equation}

\section*{Data Availability}

The training datasets and MLIP models have been publicly deposited at \url{huggingface.co/datasets/jhaens/solidacid-0_model_dataset} and \url{huggingface.co/jhaens/solidacid-0_model}.

\section*{Code Availability}

The used third-party codes \textsc{CP2K}, \textsc{ASE} and \textsc{MACE} are available at the time of publication of this work at \url{cp2k.org}, \url{ase-lib.org} and \url{github.com/acesuit/mace}. 
The high-throughput input creator used in this study is publicly available at \url{github.com/jhaens/amaceing_toolkit}.

\section*{Acknowledgments}

We thank the staff of the Compute Center of the Technische Universit{\"a}t Ilmenau and especially Mr.~Henning~Schwanbeck for providing an excellent research environment. 
This work is supported by the doctoral scholarship of the German Academic Scholarship foundation, the Carl-Zeiss-Stiftung (SustEnMat, funding code: P2023-02-008), the Th{\"u}ringer Aufbaubank (TAB) (KapMemLyse, grant no.~2024 FGR 0081 / 0082), and the European Social Fund Plus (ESF+).

\section*{Competing Interests}

The authors declare no competing interests.

\section*{Author Contributions}

J.H. conceived the idea; R.v.S. and J.H.~wrote the high-throughput workflow and performed all calculations; R.v.S. and J.H.~analysed the data; R.v.S. and J.H.~visualized all results; J.H. and R.v.S.~wrote the first draft of the manuscript; 
C.D. and J.H.~supervised the work; all authors revised and approved the manuscript.

\bibliography{literature} 

\end{document}


\title{Supplementary Information for ``Enabling Domain-Specific Atomistic Models: A Machine Learning Potential for the Solid Acid Family''}

\author{Jonas H{\"a}nseroth}
\email{jonas.haenseroth@tu-ilmenau.de}
\affiliation{Theoretical Solid State Physics, Institute of Physics, Technische Universit{\"a}t Ilmenau, 98693 Ilmenau, Germany}

\author{Rose Asuka Baroness von Stackelberg}
\affiliation{Theoretical Solid State Physics, Institute of Physics, Technische Universit{\"a}t Ilmenau, 98693 Ilmenau, Germany}

\author{Christian Dre{\ss}ler}
\affiliation{Theoretical Solid State Physics, Institute of Physics, Technische Universit{\"a}t Ilmenau, 98693 Ilmenau, Germany}

\date{\today}

\maketitle

For more information on abbreviations, please refer to the main text, where all abbreviations are defined in detail.
Abbreviations not introduced in the main text are defined here.

\tableofcontents
\newpage
\clearpage

\suppnote{Elemental composition of the training set}

\begin{figure}[ht]
    \centering
    \includegraphics{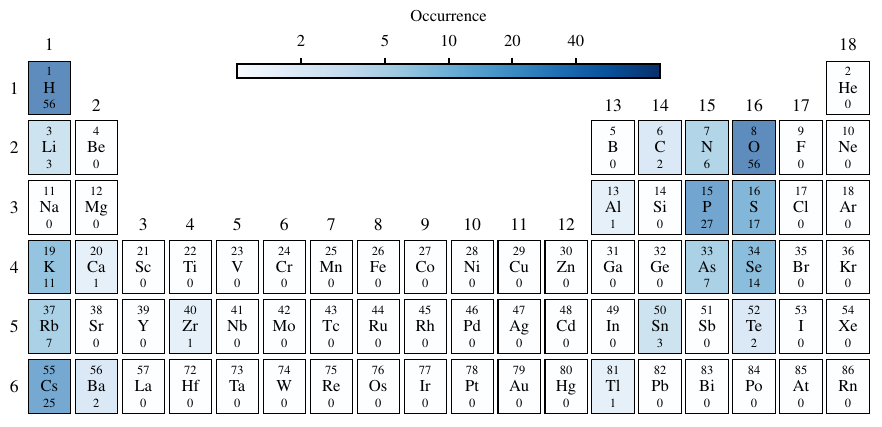}
    \caption{\textbf{Elemental diversity of the solid acid materials in the training set.}
    Periodic-table heat map summarizing the elemental composition of the 55 materials.
    Colors encode how often each element appears among the filtered structures, with the exact counts printed below the corresponding element symbols.
    The lanthanides and the seventh period are omitted because no materials in the training set contain elements from these groups.
    }
    \label{si_fig:pse}
\end{figure}

\suppnote{Distribution of the simulation cell size of the material set}

\begin{figure}[ht]
    \centering
    \includegraphics{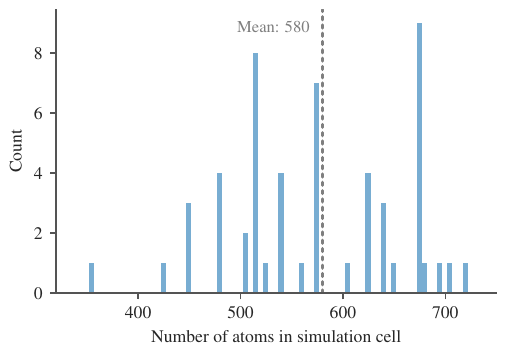}
    \caption{
    \textbf{Distribution of simulation cell sizes used for screening.}
    Histogram of the number of atoms in the simulation cells for the 55 materials inside the training set.
    The cell-size distribution reflects the use of enlarged supercells to obtain statistically meaningful proton-transport and anion-dynamics descriptors from MD simulations.
    }
    \label{si_fig:sim_cell_dims}
\end{figure}

\newpage
\suppnote{Hyper-parameter optimization}

\begin{figure}[ht]
    \centering
    \includegraphics{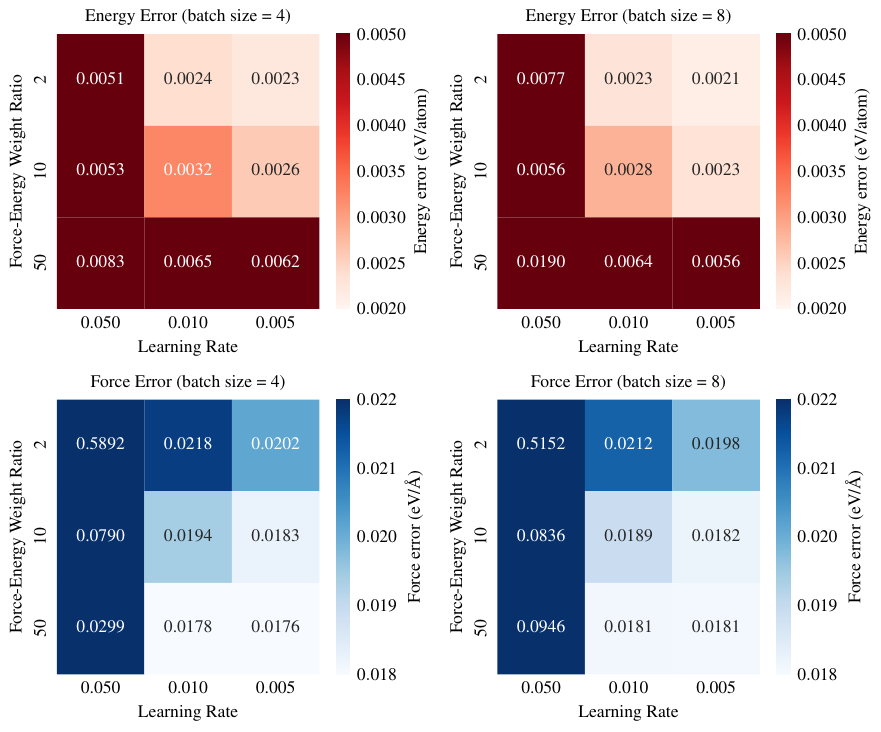}
    \caption{
    \textbf{Result of hyper-parameter optimization.}
    \textsc{MACE} models (same model size as \textsc{MACE-MP-0} (small)) were trained-from-scratch on the training set incorporating $440{,}000$ reference data points from $55$ materials with variable batch-size ($4$, $8$), learning rate ($0.005$, $0.01$, $0.05$) and force-energy loss weighting ratio ($2$, $10$, $50$) for $5$ epochs due to computational cost.
    Inside each cell the mean-absolute energy/force error averaged over all $55$ materials, in eV/atom or eV/\AA\ respectively, obtained with the corresponding model is shown.
    Darker colors correspond to larger energy/force errors.
    Hyper-parameters used for training the production models are: batch size $8$, force-energy-loss weighting ratio $10$ and learning rate $0.005$.
    }
    \label{si_fig:hyperparam_opt}
\end{figure}

\newpage
\suppnote{Error tabular overview}

\begin{figure*}[ht]
    \centering
    \includegraphics[width=\textwidth]{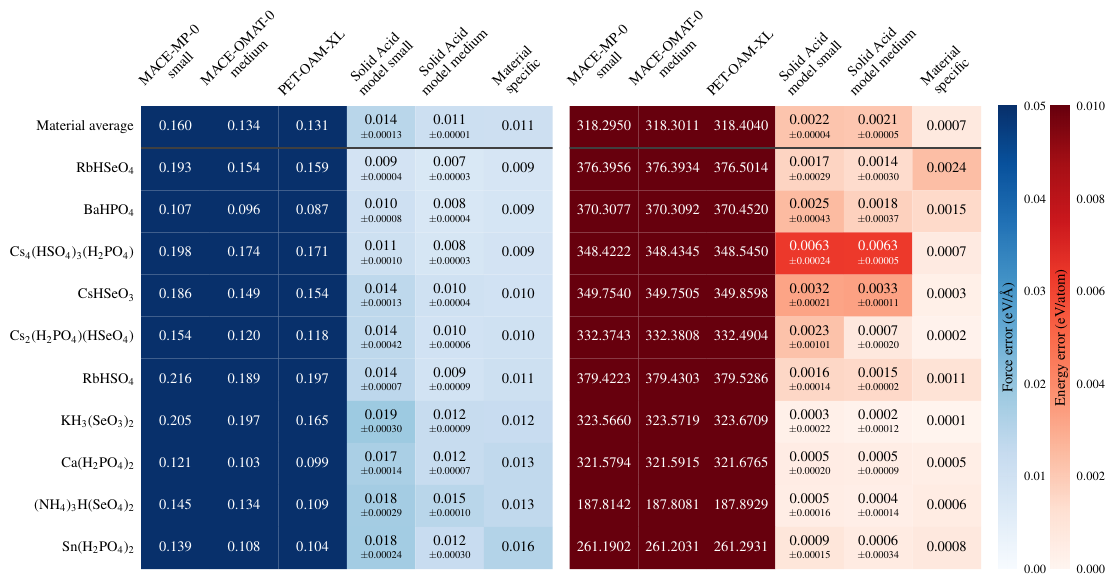}
    \caption{\textbf{Model performance evaluated on a subset of the materials.}
     Solid acid models and universal models performance on the $9$ materials in the mean absolute force and mean absolute energy errors.
     The figure shows the force and energy MAEs corresponding to main text Figure~2.
     Uncertainties on the material average are sample standard deviations of the per-seed material averages over the independently trained seed models.
     }
    \label{si_fig:overview_errors}
\end{figure*}

\newpage
\suppnote{Error tabular small model}

\begin{figure*}[ht]
    \centering
    \includegraphics[width=\textwidth]{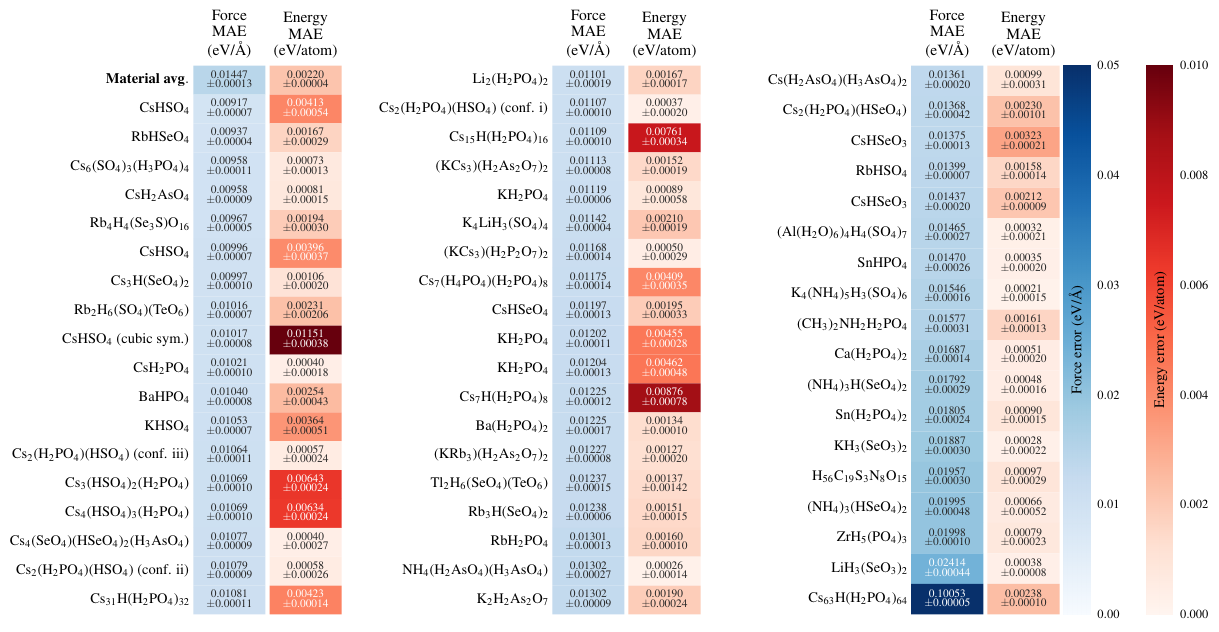}
    \caption{\textbf{Model performance evaluated on each material.}
     Local solid acid universal model (small) performance on the $55$ materials in the mean absolute force and mean absolute energy errors.
     Sorted after best force mean-absolute error.
     Uncertainties on the material average are sample standard deviations of the per-seed material averages over the independently trained seed models.
     }
    \label{si_fig:small_errors}
\end{figure*}

\newpage
\suppnote{Error tabular medium model}

\begin{figure*}[ht]
    \centering
    \includegraphics[width=\textwidth]{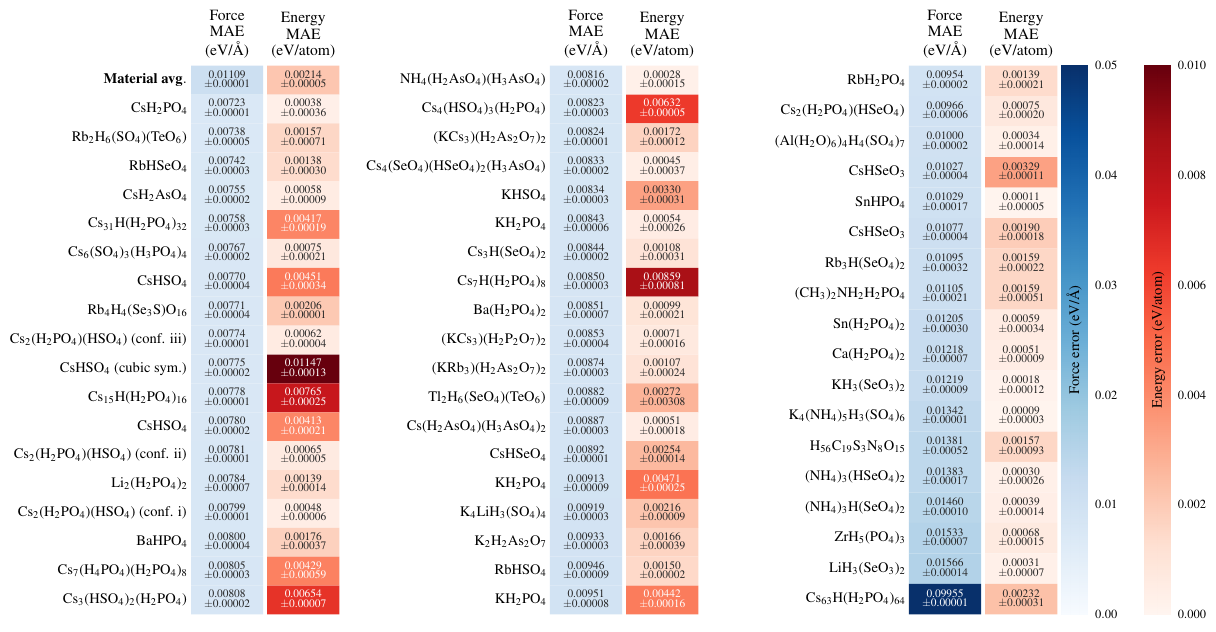}
    \caption{\textbf{Model performance evaluated on each material.}
     Local solid acid universal model (medium) performance on the $55$ materials in the mean absolute force and mean absolute energy errors.
     Sorted after best force mean-absolute error.
     Uncertainties on the material average are sample standard deviations of the per-seed material averages over the independently trained seed models.
     }
    \label{si_fig:medium_errors}
\end{figure*}

\newpage
\suppnote{Error tabular \textsc{MACE-MP-0} (small) model}

\begin{figure*}[ht]
    \centering
    \includegraphics[width=\textwidth]{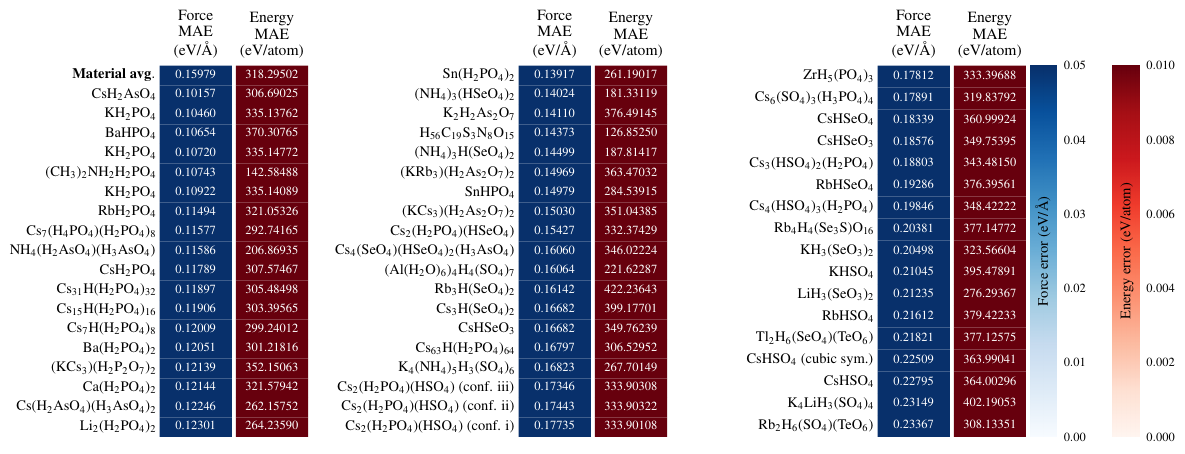}
    \caption{\textbf{Universal model performance evaluated on each material.} 
    Universal model \textsc{MACE-MP-0} (small) performance on the $55$ materials in the mean absolute force and mean absolute energy errors.}
    \label{si_fig:model_errors_mace_mp_0}
\end{figure*}

\newpage
\suppnote{Error tabular \textsc{MACE-OMAT-0} (medium) model}

\begin{figure*}[ht]
    \centering
    \includegraphics[width=\textwidth]{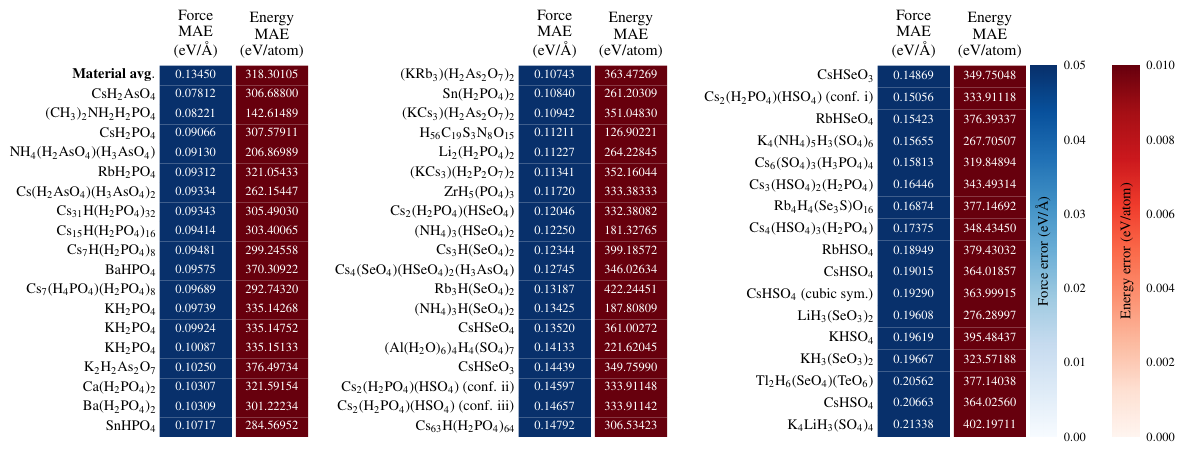}
    \caption{\textbf{Universal model performance evaluated on each material.} 
    Universal model \textsc{MACE-OMAT-0} (medium) performance on the $55$ materials in the mean absolute force and mean absolute energy errors.}
    \label{si_fig:model_errors_mace_omat_0}
\end{figure*}

\newpage
\suppnote{Error tabular \textsc{PET-OAM-XL} model}

\begin{figure*}[ht]
    \centering
    \includegraphics[width=\textwidth]{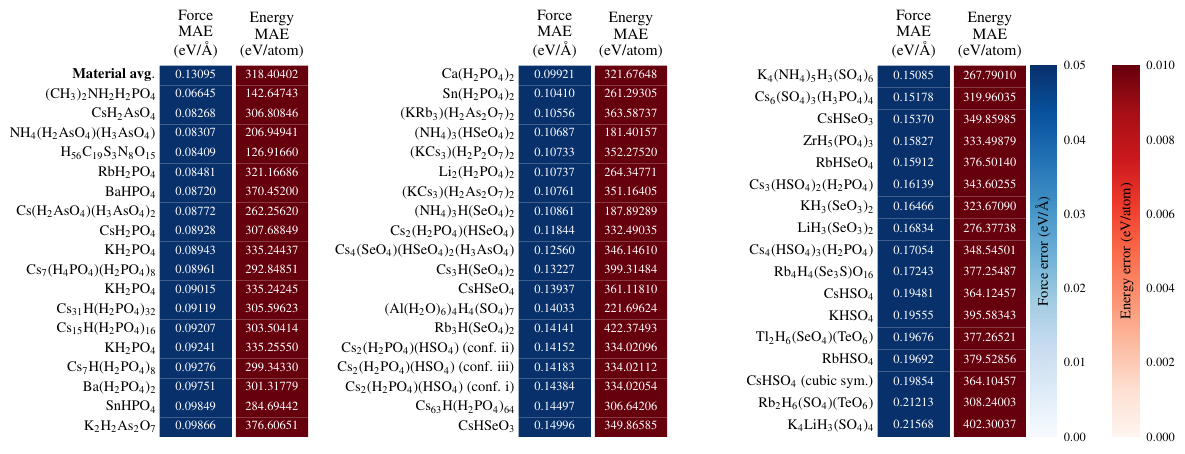}
    \caption{\textbf{Universal model performance evaluated on each material.} 
    Universal model \textsc{PET-OAM-XL} performance on the $55$ materials in the mean absolute force and mean absolute energy errors.}
    \label{si_fig:model_errors_pet_oam_xl}
\end{figure*}

\newpage
\suppnote{Error tabular sparsed out small model}

To assess the ability of the MLIPs to generalize to materials beyond those included in the training data, three additional models were trained on reduced datasets in which a subset of materials was excluded completely from training and reserved as unseen test systems. 
In each of the three cases, the unseen materials were selected randomly, but subject to additional constraints.
Specifically, materials containing elements that occur only once in the full dataset, such as Al, Ca, Zr, and Tl, were not allowed to be assigned to the unseen test set, since excluding them would remove these elements entirely from the training data. 
Furthermore, for elements or chemical motifs represented by only a small number of materials, the random selection was constrained such that not all related systems could be placed into the same unseen subset. 
This ensured that, for example, at least one material containing elements such as Li, Rb, Sn, As, N, C, or Ba remained present in the training data.

\begin{figure*}[ht]
    \centering
    \includegraphics[width=\textwidth]{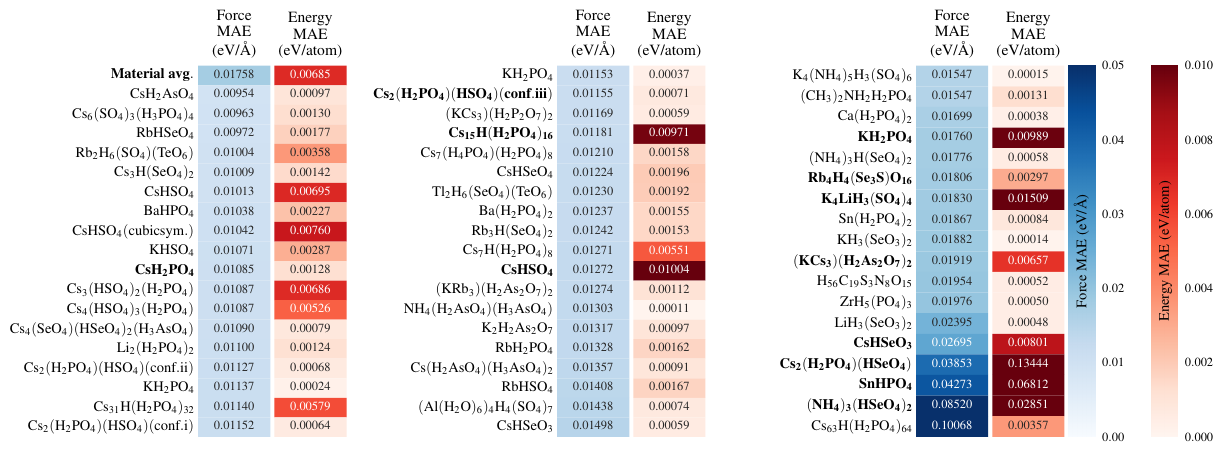}
    \caption{\textbf{Model performance evaluated on each material.}
     Local solid acid universal model (small) performance on the $55$ materials in the mean absolute force and mean absolute energy errors.
     12 randomly chosen materials (subset 1: CsHSO$_4$, CsH$_2$PO$_4$, KH$_2$PO$_4$, Cs$_2$(H$_2$PO$_4$)(HSO$_4$), Cs$_{15}$H(H$_2$PO$_4$)$_{16}$, CsHSeO$_3$, Rb$_4$H$_4$(Se$_3$S)O$_{16}$, K$_4$LiH$_3$(SO$_4$)$_4$, (KCs$_3$)(H$_2$As$_2$O$_7$)$_2$, Cs$_2$(H$_2$PO$_4$)(HSeO$_4$), SnHPO$_4$, (NH$_4$)$_3$(HSeO$_4$)$_2$)  were excluded from the training set to test the model on these unseen materials. 
     These materials are marked with a bold written chemical formula in the table.
     Sorted after best force mean-absolute error.
     }
    \label{si_fig:test_model_0_errors}
\end{figure*}

\newpage
\begin{figure*}[ht]
    \centering
    \includegraphics[width=\textwidth]{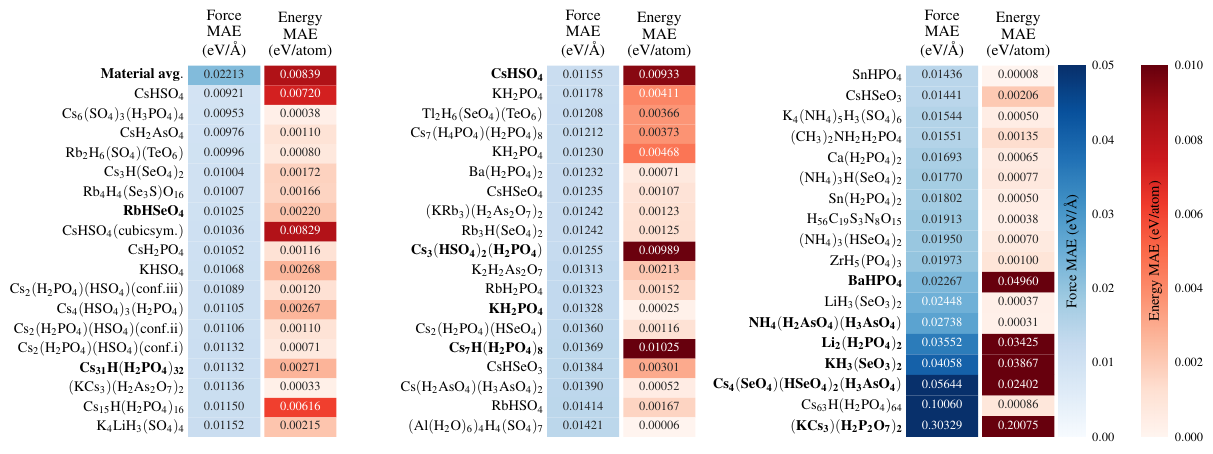}
    \caption{\textbf{Model performance evaluated on each material.}
     Local solid acid universal model (small) performance on the $55$ materials in the mean absolute force and mean absolute energy errors.
     12 randomly chosen materials (subset 2: KH$_2$PO$_4$, BaHPO$_4$, (KCs$_3$)(H$_2$P$_2$O$_7$)$_2$, RbHSeO$_4$, KH$_3$(SeO$_3$)$_2$, Cs$_{31}$H(H$_2$PO$_4$)$_{32}$, Cs$_{7}$H(H$_2$PO$_4$)$_{8}$, CsHSO$_4$, Cs$_3$(HSO$_4$)$_2$(H$_2$PO$_4$), Cs$_4$(SeO$_4$)(HSeO$_4$)$_2$(H$_3$AsO$_4$), NH$_4$(H$_2$AsO$_4$)(H$_3$AsO$_4$), Li$_2$(H$_2$PO$_4$)$_2$) were excluded from the training set to test the model on these unseen materials. 
     These materials are marked with a bold written chemical formula in the table.
     Sorted after best force mean-absolute error.
     }
    \label{si_fig:test_model_1_errors}
\end{figure*}

\newpage
\begin{figure*}[ht]
    \centering
    \includegraphics[width=\textwidth]{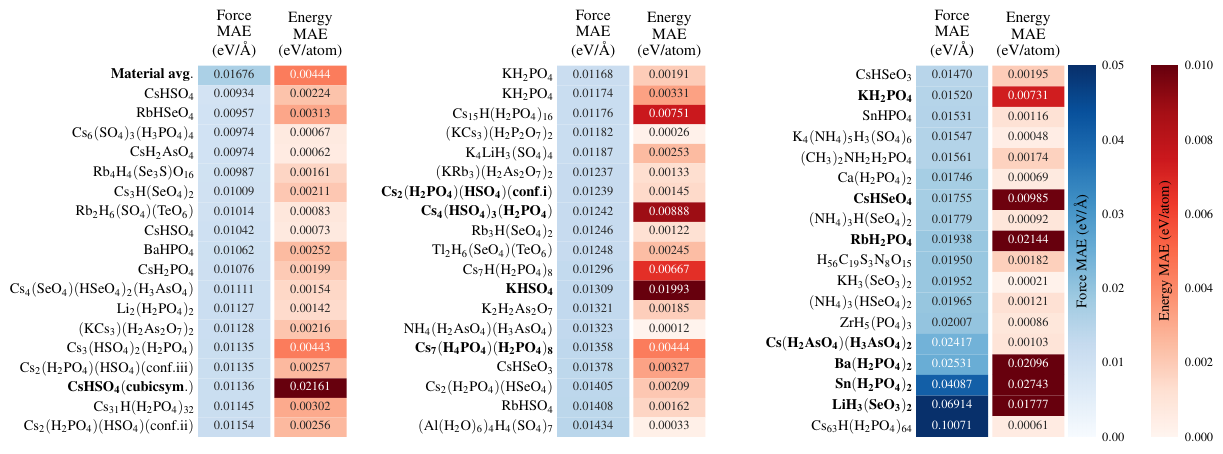}
    \caption{\textbf{Model performance evaluated on each material.}
     Local solid acid universal model (small) performance on the $55$ materials in the mean absolute force and mean absolute energy errors.
     12 randomly chosen materials (subset 3: LiH$_3$(SeO$_3$)$_2$, Ba(H$_2$PO$_4$)$_2$, KH$_2$PO$_4$, Sn(H$_2$PO$_4$)$_2$, Cs$_4$(HSO$_4$)$_3$(H$_2$PO$_4$), KHSO$_4$, Cs$_7$(H$_4$PO$_4$)(H$_2$PO$_4$)$_8$, CsHSO$_4$ (cubic symmetry), Cs$_2$(H$_2$PO$_4$)(HSO$_4$), CsHSeO$_4$, Cs(H$_2$AsO$_4$)(H$_3$AsO$_4$)$_2$, RbH$_2$PO$_4$) were excluded from the training set to test the model on these unseen materials. 
     These materials are marked with a bold written chemical formula in the table.
     Sorted after best force mean-absolute error.
     }
    \label{si_fig:test_model_2_errors}
\end{figure*}

\newpage
\suppnote{Predicting structural characteristics among the materials at GGA accuracy}
\begin{figure}[h!]
    \centering
    \includegraphics[width=0.86\textwidth]{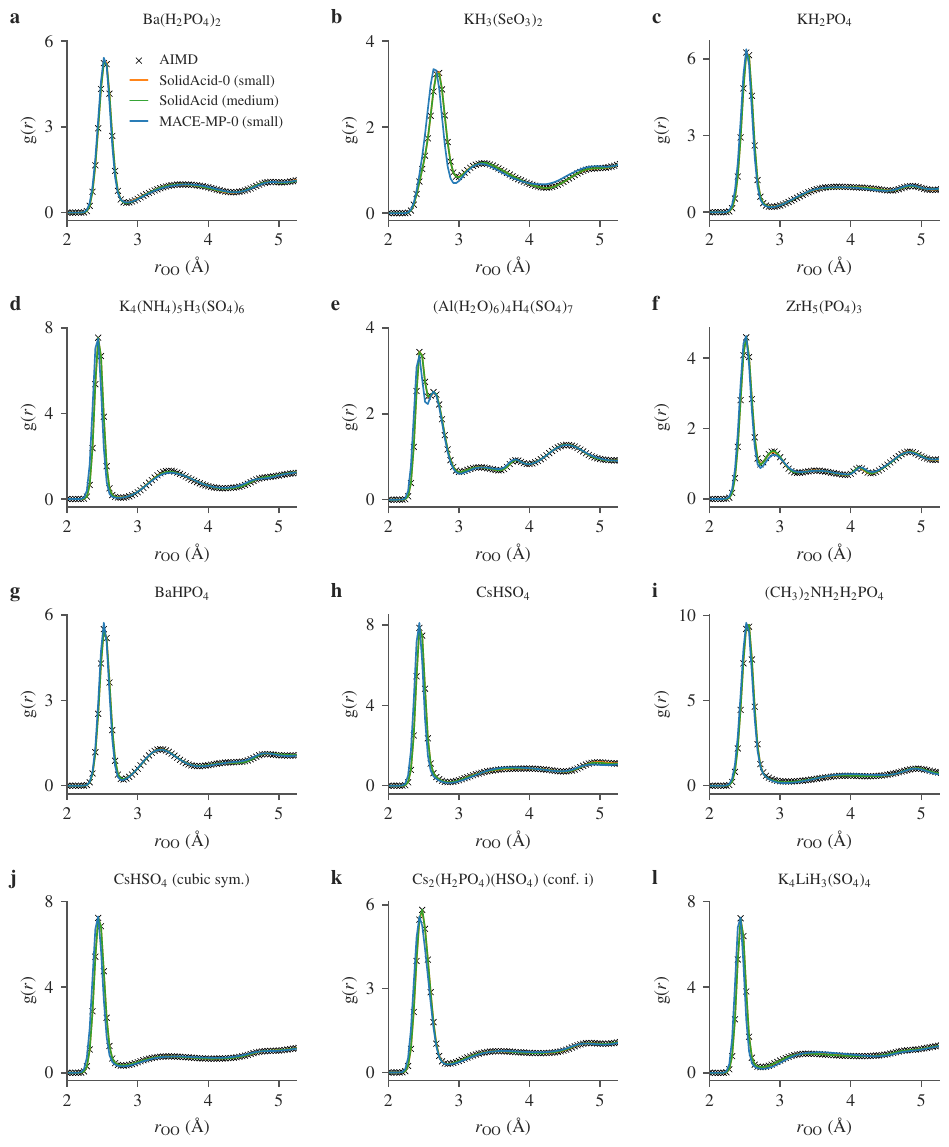}
    \caption{
    \textbf{Comparison of AIMD and local solid acid universal models O--O RDFs}.
    O--O radial distribution functions obtained from AIMD, \textsc{MACE-MP-0} (small) and the solid acid local universal \textsc{MACE} model (small \& medium) MD trajectories for twelve materials from the training set of $55$ candidates:
    \textbf{a} Ba(H$_2$PO$_4$)$_2$
    \textbf{b} KH$_3$(SeO$_3$)$_2$,
    \textbf{c} KH$_2$PO$_4$,
    \textbf{d} K$_4$(NH$_4$)$_5$H$_3$(SO$_4$)$_6$,
    \textbf{e} (Al(H$_2$O)$_6$)$_4$H$_4$(SO$_4$)$_7$,
    \textbf{f} ZrH$_5$(PO$_4$)$_3$,
    \textbf{g} BaHPO$_4$,
    \textbf{h} CsHSO$_4$,
    \textbf{i} (CH$_3$)$_2$NH$_2$H$_2$PO$_4$,
    \textbf{j} CsHSO$_4$ (cubic sym.),
    \textbf{k} Cs$_2$(H$_2$PO$_4$)(HSO$_4$) (conf. i),
    \textbf{l} K$_4$LiH$_3$(SO$_4$)$_4$. 
    }
    \label{si_fig:rdf_oo}
\end{figure}

\begin{figure}[h!]
    \centering
    \includegraphics{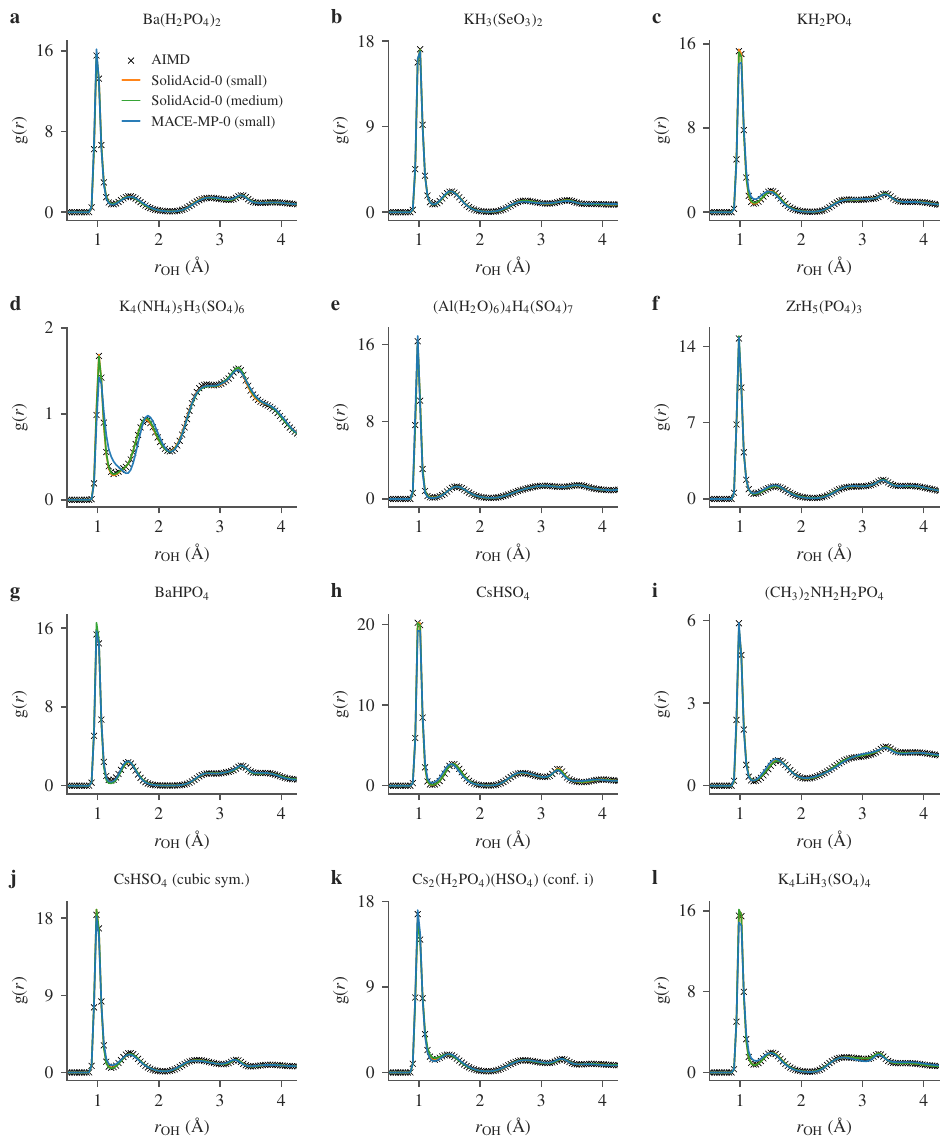}
    \caption{
    \textbf{Comparison of AIMD and local solid acid universal models O--H RDFs}.
    O--H radial distribution functions obtained from AIMD, \textsc{MACE-MP-0} (small) and the solid acid local universal \textsc{MACE} model (small \& medium) MD trajectories for twelve materials from the training set of $55$ candidates:
    \textbf{a} Ba(H$_2$PO$_4$)$_2$
    \textbf{b} KH$_3$(SeO$_3$)$_2$,
    \textbf{c} KH$_2$PO$_4$,
    \textbf{d} K$_4$(NH$_4$)$_5$H$_3$(SO$_4$)$_6$,
    \textbf{e} (Al(H$_2$O)$_6$)$_4$H$_4$(SO$_4$)$_7$,
    \textbf{f} ZrH$_5$(PO$_4$)$_3$,
    \textbf{g} BaHPO$_4$,
    \textbf{h} CsHSO$_4$,
    \textbf{i} (CH$_3$)$_2$NH$_2$H$_2$PO$_4$,
    \textbf{j} CsHSO$_4$ (cubic sym.),
    \textbf{k} Cs$_2$(H$_2$PO$_4$)(HSO$_4$) (conf. i),
    \textbf{l} K$_4$LiH$_3$(SO$_4$)$_4$.    
    }
    \label{si_fig:rdf_oh}
\end{figure}

\newpage \clearpage
\suppnote{Predicting proton diffusion and reorientation dynamics with \textsc{MACE-MP-0} (small)}
\begin{figure}[ht]
    \centering
    \includegraphics{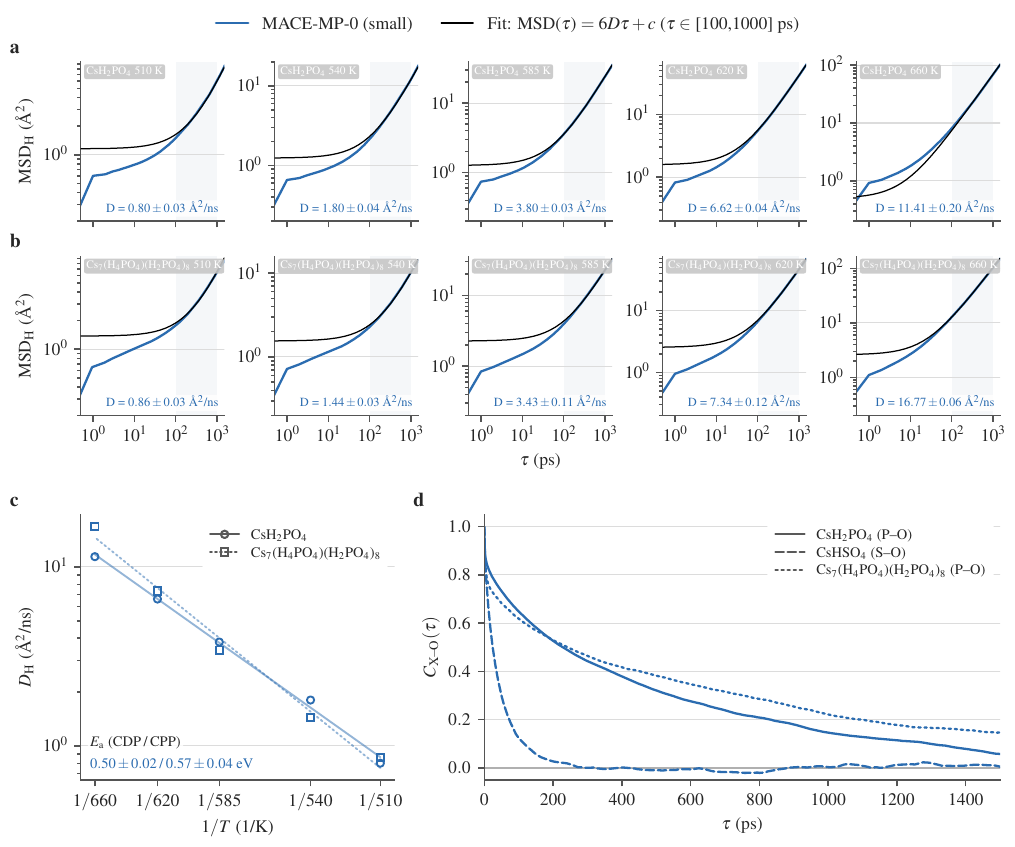}
    \caption{\textbf{Temperature and material dependence of proton diffusion and material dependence of anion reorientation.}
    Proton mean square displacements (MSDs) of CDP and CPP at $510$, $540$, $585$, $620$ and $660$~K, together with the linear fits used to determine the proton diffusion coefficients and the resulting activation energies, as well as the reorientation dynamics of CDP, CPP and CHS.
    All data were obtained from the $3$~ns MLMD trajectories calculated with the universal MLIP \textsc{MACE-MP-0} (small).
    }
    \label{sfig:mace_mp_0}
\end{figure}

\newpage
\suppnote{Predicting proton diffusion and reorientation dynamics with \textsc{MACE-OMAT-0} (medium)}
\begin{figure}[ht]
    \centering
    \includegraphics{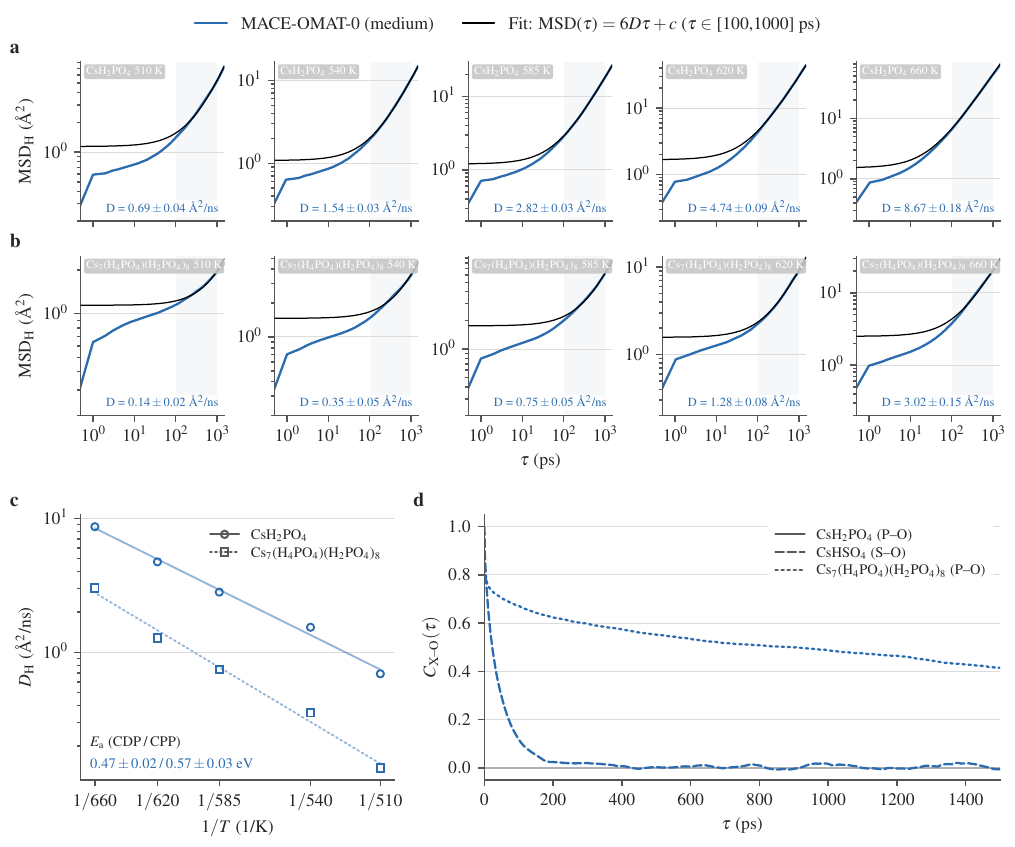}
    \caption{\textbf{Temperature and material dependence of proton diffusion and material dependence of anion reorientation.}
    Proton mean square displacements (MSDs) of CDP and CPP at $510$, $540$, $585$, $620$ and $660$~K, together with the linear fits used to determine the proton diffusion coefficients and the resulting activation energies, as well as the reorientation dynamics of CDP, CPP and CHS.
    All data were obtained from the $3$~ns MLMD trajectories calculated with the universal MLIP \textsc{MACE-OMAT-0} (medium).
    }
    \label{sfig:mace_omat_0}
\end{figure}

\newpage
\suppnote{Predicting proton diffusion and reorientation dynamics with \textsc{SolidAcid-0} (small)}
\begin{figure}[ht]
    \centering
    \includegraphics{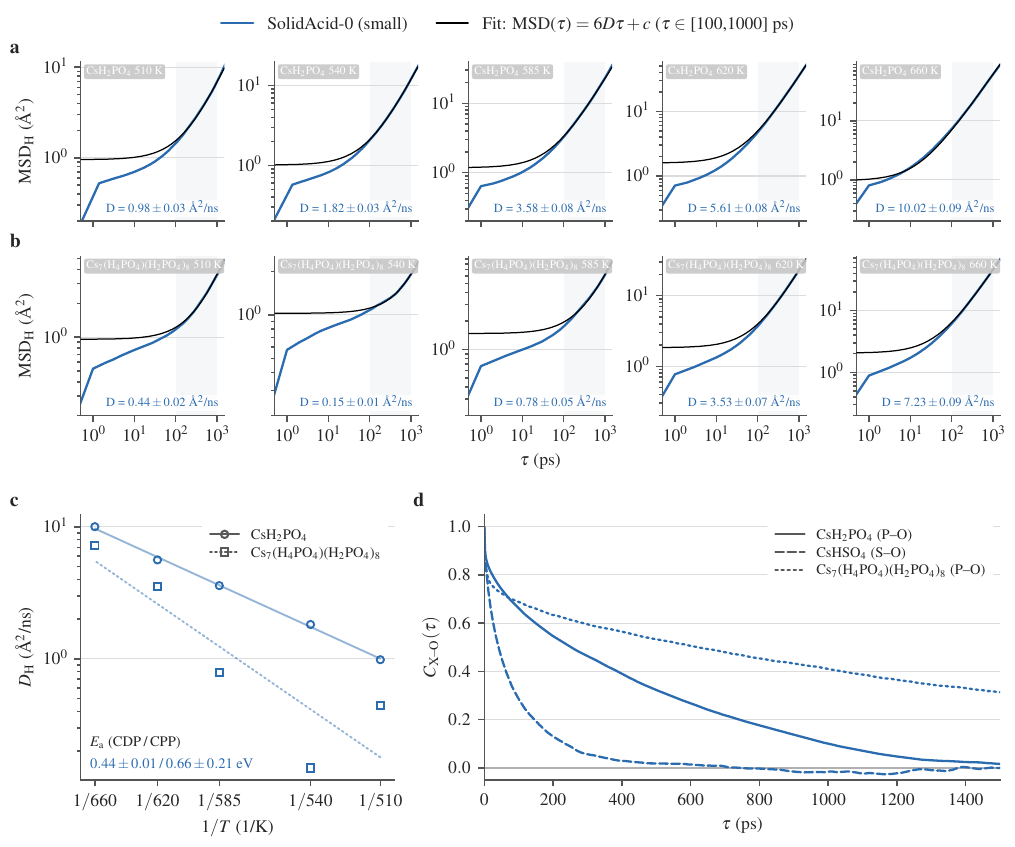}
    \caption{\textbf{Temperature and material dependence of proton diffusion and material dependence of anion reorientation.}
    Proton mean square displacements (MSDs) of CDP and CPP at $510$, $540$, $585$, $620$ and $660$~K, together with the linear fits used to determine the proton diffusion coefficients and the resulting activation energies, as well as the reorientation dynamics of CDP, CPP and CHS.
    All data were obtained from the $3$~ns MLMD trajectories calculated with the the domain specific model \textsc{SolidAcid-0} (small).
    }
    \label{sfig:solid_acid_small}
\end{figure}

\newpage
\suppnote{Predicting proton diffusion and reorientation dynamics with \textsc{SolidAcid-0} (medium)}
\begin{figure}[ht]
    \centering
    \includegraphics{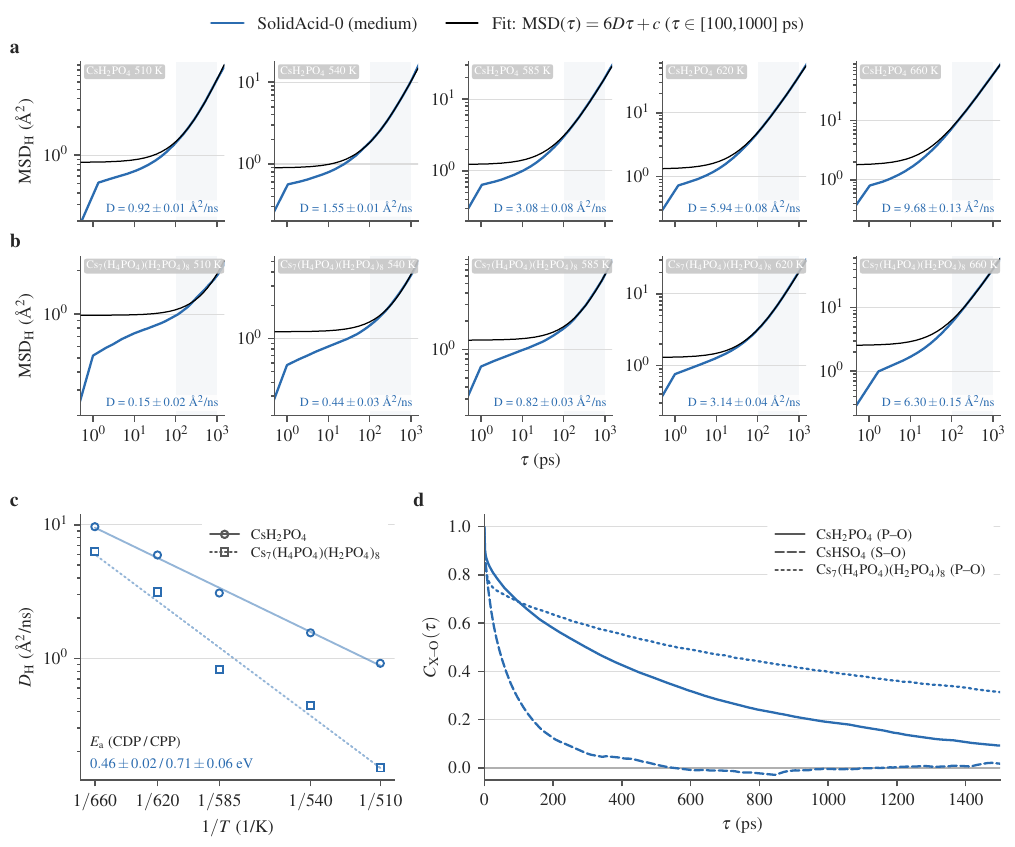}
    \caption{\textbf{Temperature and material dependence of proton diffusion and material dependence of anion reorientation.}
    Proton mean square displacements (MSDs) of CDP and CPP at $510$, $540$, $585$, $620$ and $660$~K, together with the linear fits used to determine the proton diffusion coefficients and the resulting activation energies, as well as the reorientation dynamics of CDP, CPP and CHS.
    All data were obtained from the $3$~ns MLMD trajectories calculated with the the domain specific model \textsc{SolidAcid-0} (medium).
    }
    \label{sfig:solid_acid_medium}
\end{figure}

\newpage

\newpage
\suppnote{Hybrid level dataset and fine-tuned model setup}

\begin{table}[ht]
    \centering
    \caption{
    \textbf{Hyper-parameters and model overview.}
    Hyper-parameter of the transfer-learned \textsc{SolidAcid-0} (small) \textsc{MACE} models with the reference energies used for training. 
    }
    \vspace{3mm}
    \begin{tabular}{lc@{\hspace{1em}}}
        \toprule
        \textit{Hyper-parameter} &\\
        Learning rate & $0.005$ \\
        Weight decay & $5\cdot10^{-7}$ \\
        Batch size & $5$ \\
        Force-energy-loss ratio & $10$ \\
        Number of epochs & $30$ \\
        $\max L$ & $0$ \\
        Messages & $128 \times 0e $ \\ 
        Message passing layers & $2$ \\
        Cutoff radius (\AA) & $6.0$ \\
        Angular resolution & $3$  \\
        Readout MLP irreps & $16\times0e$ \\
        Float precision & \texttt{float64} \\
        Pretrained model & \textsc{SolidAcid-0} (small) \\
        1st interaction type & \texttt{RealAgnosticDensityInteractionBlock} \\
        2nd interaction type & \texttt{RealAgnosticDensityResidualInteractionBlock} \\
        Correlation order & $3$ \\
        Optimizer & Adam \\
        Seeds & $1111$ \\
        & $2222$ \\
        & $3333$ \\
        & $4444$ \\
        & $5555$ \\
        \midrule
        \multicolumn{2}{l}{\textit{Dataset size}} \\
        Training set size & $425$  \\
        Validation set size & $75$ \\
        Test set size & $50$ \\
        \midrule
        \multicolumn{2}{l}{\textit{Reference energies} (eV)} \\
        \hspace{3em} H & -14.00875 \\
        \hspace{3em} O & -432.95278 \\
        \hspace{3em} P & -177.761933 \\
        \hspace{3em} S & -275.80771 \\
        \hspace{3em} As & -170.77918 \\
        \hspace{3em} Se & -257.000350 \\
        \hspace{3em} Cs & -552.142033 \\
    \bottomrule
    \end{tabular}
    \label{stab:tab1}
\end{table}

\newpage
\suppnote{Force and energy errors of the transfer-learned hybrid level models.}

\begin{figure*}[ht]
    \centering
    \includegraphics{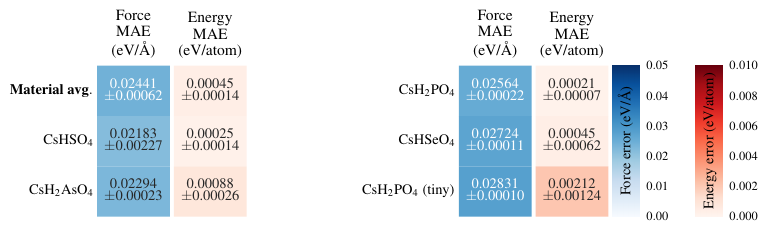}
    \caption{\textbf{Transfer-learned model performance evaluated on each material.} 
    Transfer-learned \textsc{SolidAcid-0} (tiny \& small) models performance on the $4$ materials in the mean absolute force and mean absolute energy errors.}
    \label{si_fig:model_errors_ft}
\end{figure*}

\newpage
\suppnote{Predicting structural characteristics among the materials at hybrid accuracy}
\begin{figure}[h!]
    \centering
    \includegraphics{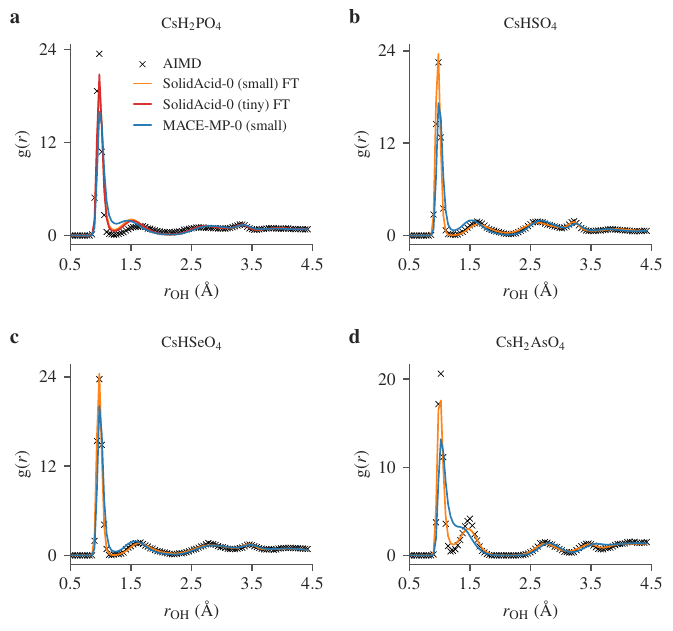}
    \caption{
    \textbf{Comparison of HSE06-AIMD and transfer-learned solid acid models O--H RDFs}.
    O--H radial distribution functions obtained from HSE06-AIMD, \textsc{MACE-MP-0} (small) and the fine-tuned solid acid local universal \textsc{MACE} model (small) MD trajectories for the four different materials in the training set:
    \textbf{a} CsH$_2$PO$_4$,
    \textbf{b} CsHSO$_4$,
    \textbf{c} CsHSeO$_3$,
    \textbf{d} CsH$_2$AsO$_4$.    
    }
    \label{si_fig:rdf_oh_ft}
\end{figure}

\begin{figure}[h!]
    \centering
    \includegraphics{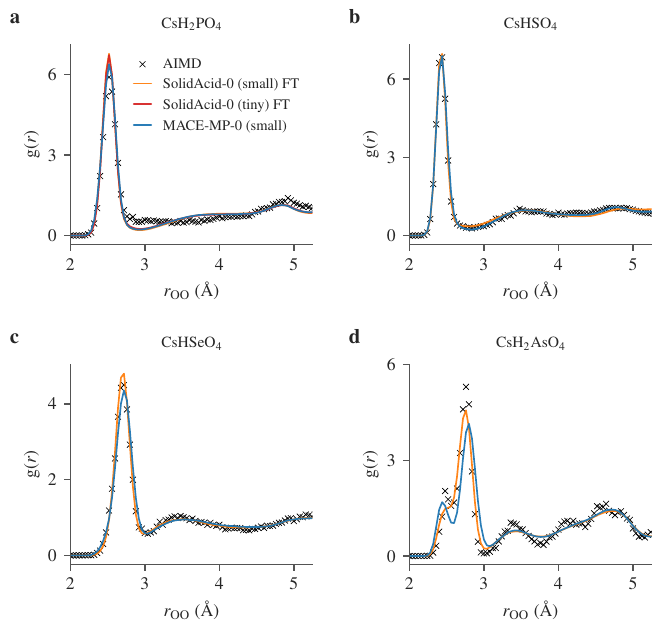}
    \caption{
    \textbf{Comparison of HSE06-AIMD and transfer-learned solid acid models O--O RDFs}.
    O--O radial distribution functions obtained from HSE06-AIMD, \textsc{MACE-MP-0} (small) and the fine-tuned solid acid local universal \textsc{MACE} model (small) MD trajectories for the four different materials in the training set:
    \textbf{a} CsH$_2$PO$_4$,
    \textbf{b} CsHSO$_4$,
    \textbf{c} CsHSeO$_3$,
    \textbf{d} CsH$_2$AsO$_4$.    
    }
    \label{si_fig:rdf_oo_ft}
\end{figure}

\newpage \clearpage
\suppnote{Predicting proton diffusion and reorientation dynamics with transfer-learned \textsc{SolidAcid-0} (small)}
\begin{figure}[ht]
    \centering
    \includegraphics{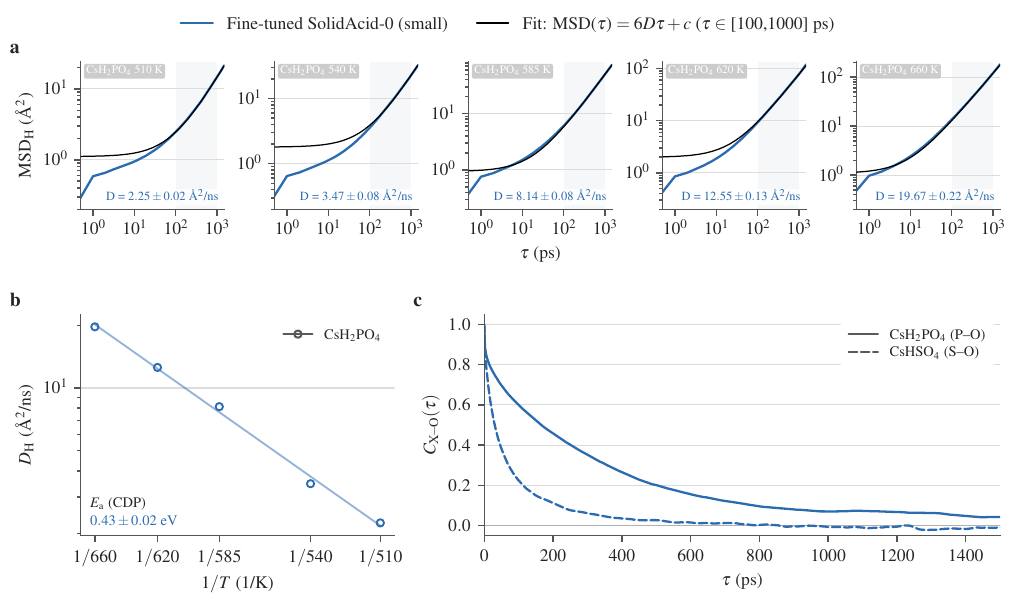}
    \caption{\textbf{Temperature and material dependence of proton diffusion and material dependence of anion reorientation.}
    Proton mean square displacements (MSDs) of CDP at $510$, $540$, $585$, $620$ and $660$~K, together with the linear fits used to determine the proton diffusion coefficients and the resulting activation energy, as well as the reorientation dynamics of CDP and CHS.
    All data were obtained from the $3$~ns MLMD trajectories calculated with the the material specific transfer-learned \textsc{SolidAcid-0} (small) model.
    }
    \label{sfig:solid_acid_small_ft}
\end{figure}

\newpage
\suppnote{Error tabular tiny model}

\begin{figure*}[ht]
    \centering
    \includegraphics[width=\textwidth]{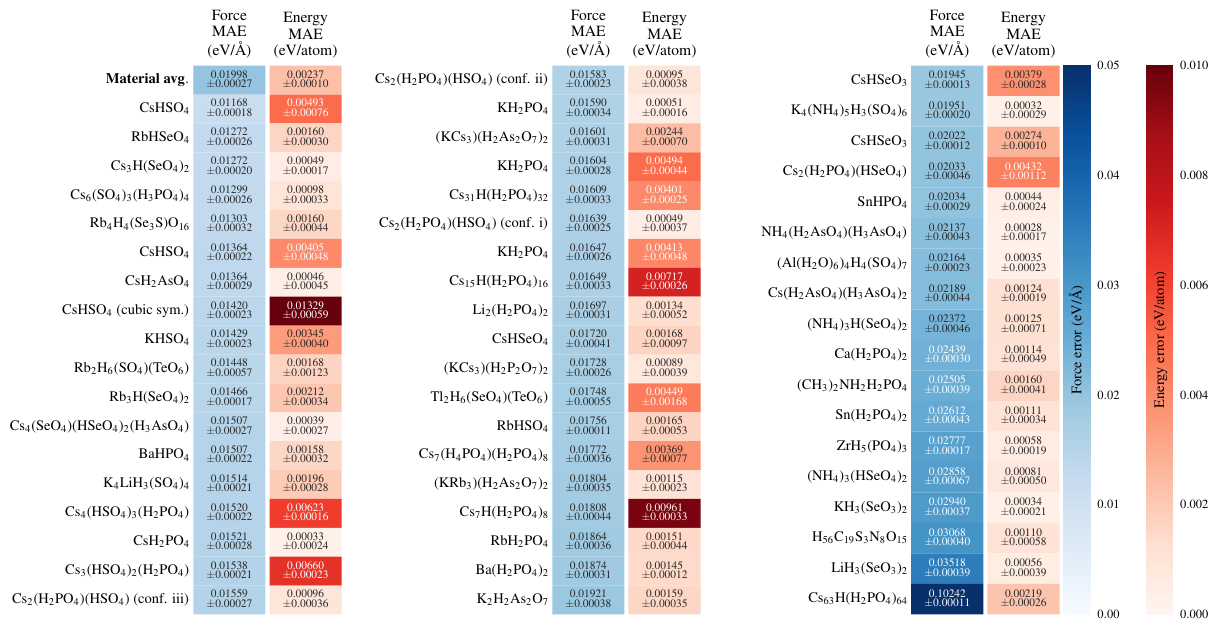}
    \caption{\textbf{Model performance evaluated on each material.}
     Local solid acid universal model (tiny) performance on the $55$ materials in the mean absolute force and mean absolute energy errors.
     Sorted after best force mean-absolute error.
     }
    \label{si_fig:tiny_errors}
\end{figure*}

\newpage
\suppnote{Predicting structural characteristics among the materials at GGA accuracy with \textsc{SolidAcid}-0 (tiny) model}
\begin{figure}[h!]
    \centering
    \includegraphics[width=0.86\textwidth]{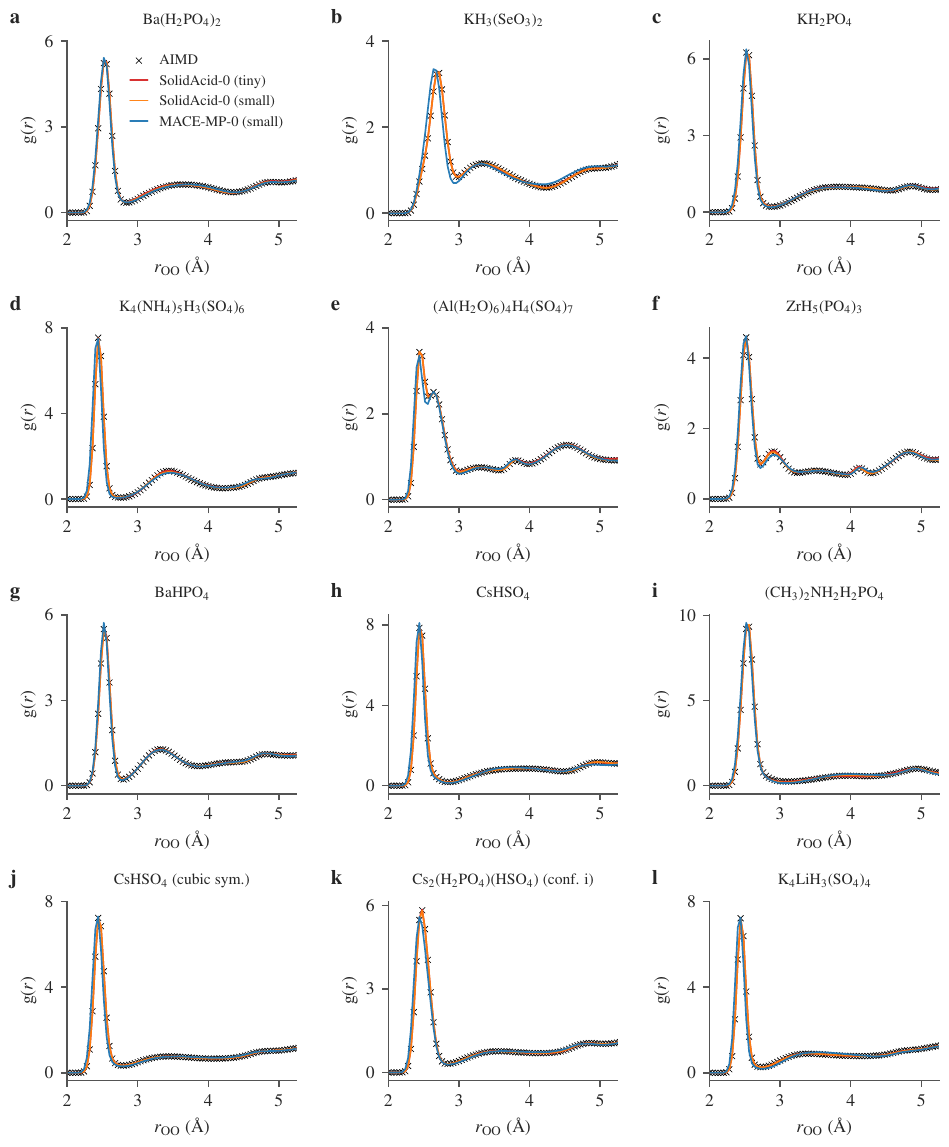}
    \caption{
    \textbf{Comparison of AIMD and local solid acid universal models O--O RDFs}.
    O--O radial distribution functions obtained from AIMD and the solid acid local universal \textsc{MACE} model (tiny \& small) MD trajectories for twelve materials from the training set of $55$ candidates:
    \textbf{a} Ba(H$_2$PO$_4$)$_2$
    \textbf{b} KH$_3$(SeO$_3$)$_2$,
    \textbf{c} KH$_2$PO$_4$,
    \textbf{d} K$_4$(NH$_4$)$_5$H$_3$(SO$_4$)$_6$,
    \textbf{e} (Al(H$_2$O)$_6$)$_4$H$_4$(SO$_4$)$_7$,
    \textbf{f} ZrH$_5$(PO$_4$)$_3$,
    \textbf{g} BaHPO$_4$,
    \textbf{h} CsHSO$_4$,
    \textbf{i} (CH$_3$)$_2$NH$_2$H$_2$PO$_4$,
    \textbf{j} CsHSO$_4$ (cubic sym.),
    \textbf{k} Cs$_2$(H$_2$PO$_4$)(HSO$_4$) (conf. i),
    \textbf{l} K$_4$LiH$_3$(SO$_4$)$_4$. 
    }
    \label{si_fig:rdf_tiny_oo}
\end{figure}

\begin{figure}[h!]
    \centering
    \includegraphics{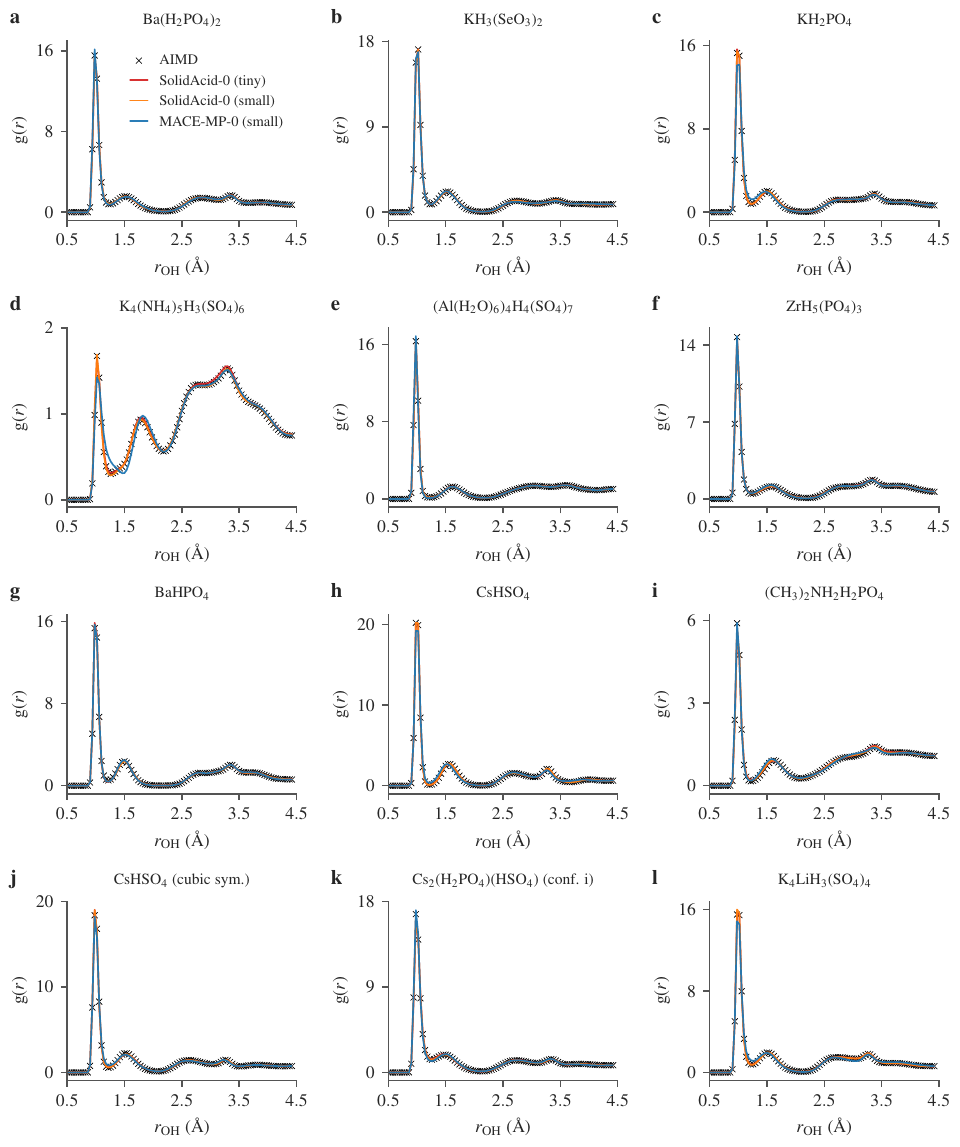}
    \caption{
    \textbf{Comparison of AIMD and local solid acid universal models O--H RDFs}.
    O--H radial distribution functions obtained from AIMD and the solid acid local universal \textsc{MACE} model (tiny \& small) MD trajectories for twelve materials from the training set of $55$ candidates:
    \textbf{a} Ba(H$_2$PO$_4$)$_2$
    \textbf{b} KH$_3$(SeO$_3$)$_2$,
    \textbf{c} KH$_2$PO$_4$,
    \textbf{d} K$_4$(NH$_4$)$_5$H$_3$(SO$_4$)$_6$,
    \textbf{e} (Al(H$_2$O)$_6$)$_4$H$_4$(SO$_4$)$_7$,
    \textbf{f} ZrH$_5$(PO$_4$)$_3$,
    \textbf{g} BaHPO$_4$,
    \textbf{h} CsHSO$_4$,
    \textbf{i} (CH$_3$)$_2$NH$_2$H$_2$PO$_4$,
    \textbf{j} CsHSO$_4$ (cubic sym.),
    \textbf{k} Cs$_2$(H$_2$PO$_4$)(HSO$_4$) (conf. i),
    \textbf{l} K$_4$LiH$_3$(SO$_4$)$_4$.    
    }
    \label{si_fig:rdf_tiny_oh}
\end{figure}

\newpage \clearpage
\suppnote{Predicting proton diffusion and reorientation dynamics with \textsc{SolidAcid-0} (tiny)}
\begin{figure}[ht]
    \centering
    \includegraphics{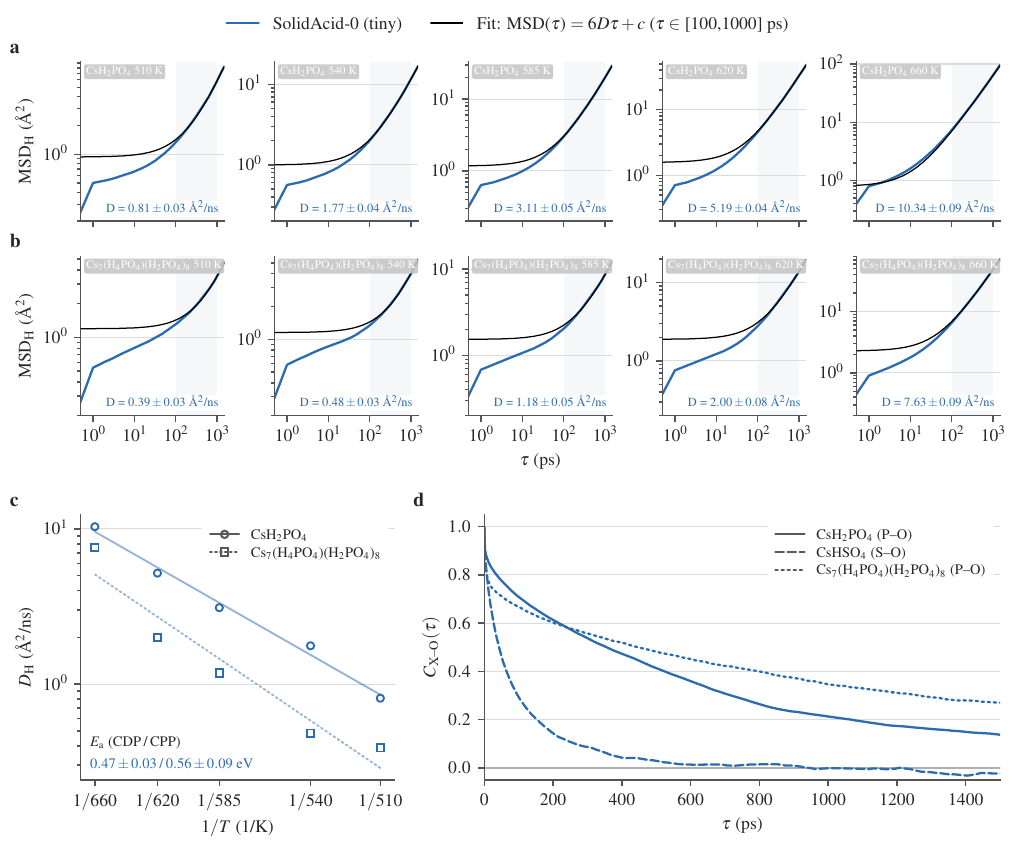}
    \caption{\textbf{Temperature and material dependence of proton diffusion and material dependence of anion reorientation.}
    Proton mean square displacements (MSDs) of CDP and CPP at $510$, $540$, $585$, $620$ and $660$~K, together with the linear fits used to determine the proton diffusion coefficients and the resulting activation energies, as well as the reorientation dynamics of CDP, CPP and CHS.
    All data were obtained from the $3$~ns MLMD trajectories calculated with the the domain specific model \textsc{SolidAcid-0} (tiny).
    }
    \label{sfig:solid_acid_tiny}
\end{figure}

\newpage 
\suppnote{Model performance evaluation}

\begin{table}[ht]
    \centering
    \caption{
    \textbf{Model performance.}
    Speed test of the \textsc{SolidAcid-0} and universal models via \textsc{ASE} molecular dynamics simulations at different precisions. 
    Benchmarked on a \textsc{NVIDIA} A100 (PCIe, $250$~W, $40$~GB).
    Compute times averaged over three independent compute runs. 
    }
    \vspace{3mm}
    \begin{tabular}{lc@{\hspace{1em}}c@{\hspace{1em}}c@{\hspace{1em}}}
            \toprule 
            Model & Parameter & $512$~atoms for $1{,}000$ steps NVE at \texttt{float32} (s) & at \texttt{float64} (s) \\
            \midrule
            \textsc{SolidAcid-0} (tiny) & $432{,}276$ & $36$ & $34$ \\
            \textsc{SolidAcid-0} (small) & $1{,}976{,}864$ & $42$ & $50$ \\
            \textsc{SolidAcid-0} (medium) & $2{,}298{,}144$ & $76$ & $101$ \\
            \textsc{MACE-MP-0} (small) \cite{mace_mp} & $3{,}847{,}696$ & $40$ & $50$ \\
            \textsc{MACE-OMAT-0} (medium) \cite{batatia2025cross} & $9{,}063{,}204$ & $77$ & $102$ \\
            \textsc{PET-OAM-XL} \cite{pet_oam} & $730{,}128{,}666$ & $759$ & -- \\
            \bottomrule
        \end{tabular}
    \label{stab:tab3}
\end{table}

\bibliography{literature}